\documentclass[12pt]{article}
\usepackage{fancyhdr}
\usepackage{mathrsfs}
\usepackage[T1]{fontenc}
\usepackage{mathpazo}
\usepackage{setspace}
\usepackage{amsfonts}
\usepackage{amssymb}
\usepackage{amsmath}
\usepackage{epsfig}
\usepackage{latexsym}
\usepackage{color}
\usepackage{graphicx}
\usepackage{nicefrac}
\usepackage[latin1]{inputenc}
\usepackage{slashed}
\usepackage{multirow}
\usepackage{comment}
\usepackage{soul}
\usepackage{hyperref}
\usepackage[nosort]{cite}
\usepackage{datetime}

\usepackage[titletoc]{appendix}

\def\hybrid{\topmargin -20pt    \oddsidemargin 0pt
        \headheight 0pt \headsep 0pt
        \textwidth 6.25in       
        \textheight 9.25in       
        \marginparwidth .875in
        \parskip 5pt plus 1pt   \jot = 1.5ex}

\hybrid

\def\baselinestretch{1.2}

\catcode`\@=11

\def\marginnote#1{}
\newcount\hour
\newcount\minute
\newtoks\amorpm
\hour=\time\divide\hour by60
\minute=\time{\multiply\hour by60 \global\advance\minute by-\hour}
\edef\standardtime{{\ifnum\hour<12 \global\amorpm={am}%
        \else\global\amorpm={pm}\advance\hour by-12 \fi
        \ifnum\hour=0 \hour=12 \fi
        \number\hour:\ifnum\minute<10 0\fi\number\minute\the\amorpm}}
\edef\militarytime{\number\hour:\ifnum\minute<10 0\fi\number\minute}

\def\draftlabel#1{{\@bsphack\if@filesw {\let\thepage\relax
   \xdef\@gtempa{\write\@auxout{\string
      \newlabel{#1}{{\@currentlabel}{\thepage}}}}}\@gtempa
   \if@nobreak \ifvmode\nobreak\fi\fi\fi\@esphack}
        \gdef\@eqnlabel{#1}}
\def\@eqnlabel{}
\def\@vacuum{}
\def\draftmarginnote#1{\marginpar{\raggedright\scriptsize\tt#1}}

\def\draft{\oddsidemargin -.5truein
        \def\@oddfoot{\sl preliminary draft \hfil
        \rm\thepage\hfil\sl\today\quad\militarytime}
        \let\@evenfoot\@oddfoot \overfullrule 3pt
        \let\label=\draftlabel
        \let\marginnote=\draftmarginnote
   \def\@eqnnum{(\theequation)\rlap{\kern\marginparsep\tt\@eqnlabel}%
\global\let\@eqnlabel\@vacuum}  }

\def\preprint{\twocolumn\sloppy\flushbottom\parindent 2em
        \leftmargini 2em\leftmarginv .5em\leftmarginvi .5em
        \oddsidemargin -.5in    \evensidemargin -.5in
        \columnsep .4in \footheight 0pt
        \textwidth 10.in        \topmargin  -.4in
        \headheight 12pt \topskip .4in
        \textheight 6.9in \footskip 0pt
        \def\@oddhead{\thepage\hfil\addtocounter{page}{1}\thepage}
        \let\@evenhead\@oddhead \def\@oddfoot{} \def\@evenfoot{} }

\def\numberbysection{\@addtoreset{equation}{section}
        \def\theequation{\thesection.\arabic{equation}}}

\def\underline#1{\relax\ifmmode\@@underline#1\else
        $\@@underline{\hbox{#1}}$\relax\fi}

\def\titlepage{\@restonecolfalse\if@twocolumn\@restonecoltrue\onecolumn
     \else \newpage \fi \thispagestyle{empty}\c@page\z@
        \def\thefootnote{\fnsymbol{footnote}} }

\def\endtitlepage{\if@restonecol\twocolumn \else \newpage \fi
        \def\thefootnote{\arabic{footnote}}
        \setcounter{footnote}{0}}  

\catcode`@=12
\relax

\def\figcap{\section*{Figure Captions\markboth
        {FIGURECAPTIONS}{FIGURECAPTIONS}}\list
        {Figure \arabic{enumi}:\hfill}{\settowidth\labelwidth{Figure
999:}
        \leftmargin\labelwidth
        \advance\leftmargin\labelsep\usecounter{enumi}}}
 \relax
\def\tablecap{\section*{Table Captions\markboth
        {TABLECAPTIONS}{TABLECAPTIONS}}\list
        {Table \arabic{enumi}:\hfill}{\settowidth\labelwidth{Table
999:}
        \leftmargin\labelwidth
        \advance\leftmargin\labelsep\usecounter{enumi}}}
 \relax
\def\reflist{\section*{References\markboth
        {REFLIST}{REFLIST}}\list
        {[\arabic{enumi}]\hfill}{\settowidth\labelwidth{[999]}
        \leftmargin\labelwidth
        \advance\leftmargin\labelsep\usecounter{enumi}}}
 \relax
\makeatletter
\newcounter{pubctr}
\def\publist{\@ifnextchar[{\@publist}{\@@publist}}
\def\@publist[#1]{\list
        {[\arabic{pubctr}]\hfill}{\settowidth\labelwidth{[999]}
        \leftmargin\labelwidth
        \advance\leftmargin\labelsep
        \@nmbrlisttrue\def\@listctr{pubctr}
        \setcounter{pubctr}{#1}\addtocounter{pubctr}{-1}}}
\def\@@publist{\list
        {[\arabic{pubctr}]\hfill}{\settowidth\labelwidth{[999]}
        \leftmargin\labelwidth
        \advance\leftmargin\labelsep
        \@nmbrlisttrue\def\@listctr{pubctr}}}
 \relax
\makeatother
\newskip\humongous \humongous=0pt plus 1000pt minus 1000pt

\newif\ifdtup

\relax

\def\be{\begin{equation}}
\def\ee{\end{equation}}
\def\ba{\begin{eqnarray}}
\def\ea{\end{eqnarray}}

\def\del{\partial}

\def\a{\alpha}

\def\b{\beta}

\def\d{\delta}

\def\l{\lambda}
\def\L{\Lambda}
\def\s{\sigma}
\def\S{\Sigma}

\def\cL{{\cal L}}

\def\no{\noindent}

\def\qq{\qquad}

\def\IR{\relax{\rm I\kern-.18em R}}

\def \ov {\over}

\def\IR{\relax{\rm I\kern-.18em R}}
\def\IL{\relax{\rm I\kern-.18em L}}

\def\inv{^{\raise.15ex\hbox{${\scriptscriptstyle -}$}\kern-.05em 1}}

\def\cL{{\cal L}}

\def\tr{{\rm tr}}
\def\Tr{{\rm Tr}}

\begin{document}

\renewcommand{\theequation}{\thesection.\arabic{equation}}
\csname @addtoreset\endcsname{equation}{section}

\newcommand{\beq}{\begin{equation}}
\newcommand{\eeq}[1]{\label{#1}\end{equation}}
\newcommand{\ber}{\begin{equation}}
\newcommand{\eer}[1]{\label{#1}\end{equation}}
\newcommand{\eqn}[1]{(\ref{#1})}
\begin{titlepage}
\begin{center}


${}$
\vskip .2 in

{\Large\bf Webs of integrable theories \\
}

\vskip 0.4in

{\large\bf George Georgiou}\ \ 
\vskip 0.15in

 {\em
Department of Nuclear and Particle Physics,\\
Faculty of Physics, National and Kapodistrian University of Athens,\\
Athens 15784, Greece\\
}
\vskip 0.12in

{\footnotesize \texttt george.georgiou@phys.uoa.gr}


\vskip .5in
\end{center}

\centerline{\bf Abstract}

\no
We present an intuitive diagrammatic representation of a new class of integrable $\s$-models. It is shown that to any given diagram corresponds  an integrable theory that couples $N$ WZW models with a certain number of each of the following four fundamental integrable models, the PCM, the YB model, both based on a group $G$,  the isotropic $\s$-model on the symmetric space $G/H$ and the YB model on the symmetric space $G/H$. To each vertex of a diagram we assign the matrix of one of the aforementioned fundamental integrable theories. Any two vertices may be connected with a number of lines having  an orientation  and   carrying an integer level $k_i$. Each of these lines is associated with an asymmetrically gauged WZW model at an arbitrary level $k_i$.  Gauge invariance of the full action is translated to level conservation at the vertices.  We also show how to immediately read from the diagrams the corresponding $\s$-model actions. The most generic of these models depends on at least $n^2+1$ parameters, where $n$ is the total number of vertices/fundamental integrable models. Finally, we discuss the case where the  level conservation at the vertices is relaxed and the case where 
the deformation matrix is not diagonal in the space of integrable models.

\vskip .4in
\noindent
\end{titlepage}
\vfill
\eject

\newpage

\tableofcontents

\noindent

\def\baselinestretch{1.2}
\baselineskip 20 pt
\noindent


\setcounter{equation}{0}
\section{Introduction }

Integrability plays a pivotal role in  obtaining exact results in quantum field theory (QFT). One of the  most studied examples in which integrability
was greatly exploited is that of $\mathcal N=4$ SYM, the maximally supersymmetric gauge theory in four spacetime dimensions. 
Employing a variety of integrability-based techniques ranging from the asymptotic Bethe ansatz \cite{Staudacher:2004tk} and the thermodynamic Bethe ansatz \cite{Ambjorn:2005wa} to the Y-system \cite{Gromov:2009tv},  the planar anomalous dimensions of gauge invariant operators was determined essentially for all values of the 't Hooft coupling $\l=g_{YM}^2 N$.  Further developments on integrability and  the AdS/CFT correspondence can be found in \cite{Beisert:2010jr} and references therein.

Integrable non-linear $\s$-models play an instrumental role in the context of gauge/ gravity dualities. This happens because, thanks to the duality, the strongly coupled dynamics of gauge theory can be translated to the 
weakly coupled dynamics of an integrable two-dimensional non-linear $\s$-model. The prototypical example of such an integrable  $\s$-model is the principal chiral model (PCM) based on a semi-simple group $G$, with or without  a Wess-Zumino (WZ) term. In \cite{Klimcik:2002zj,Klimcik:2008eq,Klimcik:2014} it was shown that the PCM based on a semi-simple group $G$ admits an integrable deformation depending on an additional  continuous parameter. These integrable models are called Yang-Baxter (YB) models and for the case of symmetric and semi-symmetric spaces they were studied in  \cite{Delduc:2013fga,Delduc:2013qra,Arutyunov:2013ega}.
There are also two parameter integrable deformations of the PCM. These are the YB $\s$-model with a WZWN term \cite{Delduc:2014uaa} and the bi-YB model \cite{Klimcik:2014}. Furthermore, integrable deformations of the PCM with three or more parameters were studied in \cite{Delduc:2017fib}.
It is a remarkable fact that all these models can be put under the unifying description of the so-called $\cal E$-models \cite{Klimcik:2015gba,Klimcik:2017ken}.

Recently, the systematic construction of a large class of integrable two-dimensional field theories based on group, symmetric and semi-symmetric spaces  and having an explicit Lagrangian formulation was deployed in a series of papers \cite{Sfetsos:2013wia,Georgiou:2016urf,Georgiou:2017jfi,Georgiou:2018hpd,Georgiou:2018gpe,Hollowood:2014rla,Hollowood:2014qma,Driezen:2019ykp}. 
These models may contain several couplings, for small values of which they take the form of  one or more WZW models \cite{Witten:1983ar}  perturbed by current bi-linears. 
Following their construction, the quantum properties of these theories were studied in great detail in 
\cite{Georgiou:2015nka,Georgiou:2016iom,Georgiou:2016zyo,Georgiou:2017oly,Itsios:2014lca,Georgiou:2017aei}. 
In this context many observables of these theories, including their $\beta$-functions\cite{Kutasov:1989dt,Gerganov:2000mt,Itsios:2014lca,Sfetsos:2014jfa,LeClair:2001yp,Appadu:2015nfa}, anomalous dimensions of currents and primary operators \cite{Georgiou:2015nka,Georgiou:2016iom,Georgiou:2016zyo,Georgiou:2019jcf,Georgiou:2019aon} and three-point correlators of currents and/or primary fields \cite{Georgiou:2015nka,Georgiou:2019aon} were computed as {\it exact} functions of the deformation parameters.
Subsequently, the Zamolodchikov's C-function \cite{Zamolodchikov:1986gt} of these models were calculated also
as {\it exact} functions of the deformation parameters \cite{Georgiou:2018vbb,Sagkrioti:2018abh}.\footnote{These results although exact in the deformation parameters provide only the leading contribution in the $1/k$-expansion. More recently, the subleading terms in $1/k$-expansion were obtained 
for the $\beta$-functions in \cite{Georgiou:2019nbz,Hoare:2019mcc} and for the C-function and the anomalous dimensions of the operators perturbing the CFT in the cases of group and coset spaces in \cite{Georgiou:2019nbz} .} 

To get these exact results for the aforementioned observables a variety of complementary methods were employed. One way \cite{Georgiou:2016iom,Georgiou:2015nka,Georgiou:2016zyo} to obtain exact expressions for the anomalous dimensions of currents and primary operators, as well as
for the three-point correlators involving currents and primaries 
was to combine low order perturbation theory around the conformal point  with certain non-perturbative  symmetries  \cite{Kutasov:1989aw,Georgiou:2018hpd,Georgiou:2018gpe,Georgiou:2016zyo} which these theories generically exhibit in the space of couplings.  Another method 
developed was based on the geometry in the space of couplings \cite{Georgiou:2019jcf}. This method makes no use of perturbation theory and  allows, in principle, the calculation of the anomalous dimensions of composite operators made from an arbitrary number of currents. The essence of the method relies on the ability to construct  the all-loop effective action of these models\cite{Georgiou:2019jcf}.
Even more recently, yet another method for calculating exact results in this class of models was initiated in \cite{Georgiou:2019aon}. The method consists of expanding the known all-loop effective actions of the theories around 
the unit group element and keeping only a few leading terms in the expansion. 
The advantage of this method  is that one ends up performing perturbative calculations around a free field theory and not around the conformal point, which is a much easier task. In addition, all 
deformation effects are captured by the couplings of the interaction vertices. Subsequently, the applicability of this method
to the case of deformed coset CFTs was demonstrated in
\cite{Georgiou:2020bpx}. 

Let us mention that the main virtue of the models constructed in \cite{Georgiou:2017jfi,Georgiou:2018hpd,Georgiou:2018gpe} for deformations
based on current algebras  and in \cite{Sfetsos:2017sep} for deformations  of coset CFTs,  compared to the prototype single $\l$-deformed model of \cite{Sfetsos:2013wia} (for the group $SU(2)$ the $\l$-deformed model was found earlier in \cite{Balog:1993es}) is that the RG flows of the former have a rich structure consisting of several fixed points, with  different CFTs
sitting at different fixed points. It remains an open problem to fully classify these CFTs according to their symmetry groups. 
In \cite{Georgiou:2020eoo}, this goal was achieved for a generalisation of the cyclic $\l$-deformed models of \cite{Georgiou:2017oly} in which arbitrary different levels for the WZW models were allowed.

\no
In a parallel development, an interesting relation between $\l$-deformations and $\eta$-deformations for group and coset spaces was uncovered in  \cite{Vicedo:2015pna,Hoare:2015gda}, \cite{Sfetsos:2015nya,Klimcik:2015gba,Klimcik:2016rov,Hoare:2018ebg}. In particular, the  $\l$-deformed models are related to the $\eta$-deformed models via Poisson-Lie T-duality \footnote{Poisson-Lie T-duality has been introduced for group spaces in  \cite{KS95a} and extended to coset spaces in
\cite{Sfetsos:1999zm}.} and appropriate analytic continuations.
Finally,  D-branes regarded as  boundary configurations  preserving integrability were introduced in the context of $\lambda$-deformations in
\cite{Driezen:2018glg}.

\no
The plan of the paper is as follows.
In section 2, we will construct the $\s$-model actions
of a general class of integrable models
that couple $N$ WZW models with an arbitrary number of the following four fundamental  theories, that is $n_1$ different copies of  the PCM, $n_2$ different copies of the YB model, both based on a group $G$,   $n_3$ different copies of the   the isotropic $\s$-model on the symmetric coset space $G/H$ and $n_4$ different copies of the YB model on the symmetric space $G/H$. The coupling is achieved by gauging 
the left global symmetry of the aforementioned fundamental integrable models and connecting them with asymmetrically gauged WZW models. The latter depend on both the gauge fields of the fundamental integrable theories which 
they connect.
 In this way, webs of integrable theories are obtained. We show that a diagrammatic 
representation of these webs is possible. The virtue of this diagrammatic 
representation is that one can, at the back of the envelope, draw any diagram and directly write down from it the corresponding integrable theory. 
The model corresponding to a diagram in which all possible kinds of lines and vertices are present
depends on at least $n^2+1+n_2+ n_4$ parameters, where $n$ is the total number of vertices/fundamental integrable models. For small values of the deformation parameters the $\sigma$ models obtained after integrating out the gauge fields are $N$ couples WZW models perturbed by current-current interaction terms of a specific form (see \eqref{small-lam}).

In section 3, we will prove that the the theories constructed in section 2 are indeed classically integrable by finding  the corresponding Lax pairs  for $n$ distinct combinations of the equations of motion. The remaining $N-n$ equations of motion take the form of covariantly free combinations of currents. Each of these covariantly free equations give rise to an infinite tower of local conserved  charges which supplement the ones obtained from the Lax pairs, in the usual way. As a result, one gets as many infinite towers of conserved 
charges as the degrees of freedom of the theories which proves that our theories are integrable.

In section 4, we will consider two more general situations. In the first one we focus on the case in which  the deformation matrix is not diagonal in the space of  the fundamental theories, in distinction with the models of section 2.
In the second, we examine the case in which,  although the deformation matrix
 is diagonal in the space of  the fundamental theories,   level conservation at the vertices is relaxed. In both cases we were able to prove integrability
only when all the deformation matrices are proportional to the identity in the group space, that is when only when the theories we couple are all of the PCM-type.
Finally, in section 5 we will present our conclusions.

\section{Coupling integrable theories }
In this section we will construct the effective actions of our models and establish their diagrammatic representation.  In section 3, we will derive the corresponding equations of motion and prove that the theories presented in the present section  are classically integrable.
\subsection{Constructing the models and their diagrammatic representation}
Our starting point is to consider the sum of $n$ integrable models based on group elements $\tilde g_i$, $i=1,2,\dots ,n$ each of which has a left global symmetry $\tilde g_i\to \L_i^{-1} \tilde g_i$,  which will be 
eventually gauged. Thus, we start from the action
\be
\label{bb}
\begin{split}
& S_{E_i}(\tilde g_i)= 
-{1\ov \pi}\int d^2\s\ \big(\tilde g_i^{-1}\del_+\tilde g_i\big)_a  E_{ij}^{ab} \big(\tilde g_j^{-1} \del_-\tilde g_j\big)_b\ ,\qquad E_{ij}^{ab}=\d_{ij} E_i^{ab}\,
\end{split}
\ee
where the indices $i,j=1,2,\dots ,n$ enumerate the different integrable models while the indices $a,b=1,2,\dots, \dim(G)$ denote group indices. Furthermore, although in most of this paper we will assume that 
the matrix $E_{ij}\sim \d_{ij}$ in all algebraic manipulations we will keep its most general non-diagonal in the space of models form, in anticipation of the analysis of section 4.1.
The integrable models appearing in \eqref{bb} will be the basic building blocks of our construction and will be called the fundamental integrable models. In the sum \eqref{bb} there can be $n_1$ different copies of the PCM, 
$n_2$ different copies of the YB model both based on the same semi-simple group $G$,  $n_3$ different copies of  the isotropic $\s$-model on the symmetric coset space $G/H$ and $n_4$ different copies of the YB model on the symmetric space $G/H$, \,\, 
with $n_1+n_2+n_3+n_4=n$. \\
The corresponding $E_i$ matrices 
acquire the following forms, namely $E_i^{ab}=E_i \,\d^{ab}$ for the PCM,  $E_i={1\ov t_i}(1-\eta_i {\cal R}_i)^{-1}$  for the YB, $E_i^{ab}=diag(E_i^{g/h}\d^{ab},0_h)$ for the isotropic symmetric space $G/H$ and $E_i=diag({1\ov t_i}(1-\eta_i {\cal R}_i)^{-1}|_{g/h},0_h)$ for the YB on 
the symmetric space $G/H$. In the last case ${\cal R}_i$ is an antisymmetric matrix of dimension $dim(G)-dim(H)$ which one can think of as being the projection to the coset $G/H$ of an ${\cal R}$-matrix obeying the modified Yang-Baxter equation \cite{Sfetsos:2015nya}.\footnote{For the YB theories the group indices $a,b$ have been suppressed in the corresponding expressions for $E_i$.} In addition, ${\cal R}_i$ should be such that it obeys the condition \eqref{cond}. At this point let us mention that the constraint \eqref{cond} is a stringent one. In fact, there are cases where the  constraint is satisfied. These include the ${SU(2)\ov U(1)}$ coset space \cite{Sfetsos:2015nya}, as well as $\sigma$-models on $CP^n$ with $n>1$\cite{Demulder:2020dlo}. However, at the level of the classical $\sigma$-model one of the deformation parameters can be eliminated by a suitable redefinition of the parameters. For the case of the $\sigma$-models on $CP^n$ \cite{Demulder:2020dlo}, this redefinition is given by equation (3.11) (see also the discussion following equation (4.10), as well as point 3 on page 3 of the same work).\footnote{The arguments of  \cite{Demulder:2020dlo} also apply to our case since the two-parameter
deformed models presented in \cite{Demulder:2020dlo}  are related to the two-parameter $\lambda$-deformed models through a Poisson-Lie T-duality and an analytic continuation.}
Therefore, it remains to be seen if there are any non-trivial examples based on this type of deformation.\footnote{We thank K. Siampos for useful discussions on this point.}


The question we would like to answer in this section is the following. Is it possible to  connect the aforementioned fundamental integrable models appearing in \eqref{bb} in such a way that the resulting $\s$-model is also integrable?
The answer to this question is affirmative. The first step to achieve this  goal is to gauge the left global symmetry of  \eqref{bb} mentioned above. As a result the action \eqref{bb} becomes
\be
\label{bc}
\begin{split}
& S_{E_i}(\tilde g_i,A_\pm^{(i)})= 
-{1\ov \pi}\int d^2\s\ \big(\tilde g_i^{-1}\tilde D_+\tilde g_i\big)_a  E_{ij}^{ab} \big(\tilde g_j^{-1} \tilde D_-\tilde g_j\big)_b\ ,\qquad E_{ij}^{ab}=\d_{ij} E_i^{ab}\,
\end{split}
\ee
where the covariant derivatives are defined as
$\tilde D_\pm \tilde g_i= (\del_\pm -A_\pm^{(i)}) \tilde g_i$.
The second step is realised by connecting the gauged models in \eqref{bc}
with asymmetrically gauged WZW models at arbitrary integer levels. To be more precise consider the asymmetrically gauged WZW model \cite{Witten:1991mm}
\be\label{asWZW}
\begin{split}
&  S_{k_{ij}^{(l_{ij})}}(g_{ij}^{(l_{ij})},A_-^{(i)},A_+^{(j)}) = S_{k_{ij}^{(l_{ij})}}(g_{ij}^{(l_{ij})})
+{k_{ij}^{(l_{ij})}\ov \pi} \int d^2\s \ \Tr \big(A_-^{(i)} J_{+\,ij}^{(l_{ij})}   - A_+^{(j)} J_{-\,ij}^{(l_{ij})}
\\
& \qq\qq\qq + A_-^{(i)} g_{ij}^{(l_{ij})}A_+^{(j)} \big(g_{ij}^{(l_{ij})}\big)^{-1}-{1\ov 2 }A_-^{(i)}A_+^{(i)}-{1\ov 2 }A_-^{(j)}A_+^{(j)}\big) \ ,
\end{split}
\ee
where we have defined the currents \footnote{Regarding the WZW action and the Polyakov -Wiegmann identity we follow the conventions of \cite{Georgiou:2016urf,Georgiou:2017jfi}.}
\be\label{defj}
J_{+\,ij}^{(l_{ij})}=J_+(g_{ij}^{(l_{ij})})=\partial_+g_{ij}^{(l_{ij})} \big(g_{ij}^{(l_{ij})}\big)^{-1}\, , \qquad J_{-\,ij}^{(l_{ij})}=J_-(g_{ij}^{(l_{ij})})=\big(g_{ij}^{(l_{ij})}\big)^{-1}\partial_-g_{ij}^{(l_{ij})},
\ee
and where $S_{k_{ij}^{(l_{ij})}}(g_{ij}^{(l_{ij})})$ is the WZW model at level $k_{ij}^{(l_{ij})}$.
The notation in \eqref{asWZW} and \eqref{defj} should be self-explanatory. The asymmetrically gauged WZW functional depends on the group element $g_{ij}^{(l_{ij})} $ and connects the fundamental integrable model at site $i$ to that at site $j$ since it depends also on $A_-^{(i)}$ and $A_+^{(j)}$.
The corresponding WZW level is denoted by $k_{ij}^{(l_{ij})}$. The superscript $l_{ij}$ counts how many different gauged WZW models connecting site $i$ to  site $j$ one has. In the case where there is just one such model the superscript $l_{ij}$ is superfluous and can be omitted (see, for example, figure 3). Furthermore, due to the asymmetry of the gauging one can assign a direction to the WZW model, and as a consequence to the flow of the level $k_{ij}^{(l_{ij})}$, which we choose to be from the site $i$ to the site $j$. Notice that $k_{ij}^{(l_{ij})}$ is generically different from $k_{ji}^{(l_{ji})}$ due to the asymmetry mentioned above, the former connects sites $i$ and $j$ having direction from $i$ to $j$ while the latter connects the same sites but with opposite direction. An important comment is in order. In the case where $i\equiv j$ \eqref{asWZW} becomes the usual vectorially gauged WZW model at level $k_{ii}^{(l_{ii})}$.

 The group elements of the asymmetrically gauged WZW models have the following transformations $g_{ij}^{(l_{ij})}\to \L_i^{-1} g_{ij}^{(l_{ij})}\L_j$. 
 Needless to say that as it stands the action \eqref{asWZW} is not gauge invariant. Its variation under the infinitesimal form of the gauge transformations
\be\label{trans}
  \d g_{ij}^{(l_{ij})} = g_{ij}^{(l_{ij})} u_j - u_i g_{ij}^{(l_{ij})} \ ,\qq
\d A_\pm^{(i)} =-\del_\pm u_i + [A_\pm^{(i)}, u_i]\ ,\qq
\d A_\pm^{(j)} =-\del_\pm u_j + [A_\pm^{(j)}, u_j]\ ,
\ee
is given by 
\be
\d S_{k_{ij}^{(l_{ij})}}(g_{ij}^{(l_{ij})},A_-^{(i)},A_+^{(j)})  = {k_{ij}^{(l_{ij})}\ov 2\pi} \int d^2\s \
\Tr\big[ (A_+^{(i)} \del_- u_i - A_-^{(i)} \del_+ u_i) - (A_+^{(j)}  \del_- u_j - A_-^{(j)}  \del_+ u_j)\big]\ .
\label{dS}
\ee
Notice that in the special case where $i\equiv j$,\, $\d S_{k_{ii}^{(l_{ii})}}(g_{ii}^{(l_{ii})},A_-^{(i)},A_+^{(i)})=0$.

Consider now the complete action
\be\label{full-act}
S_t= S_{E_i}(\tilde g_i,A_\pm^{(i)})+\sum_{i,j}\sum_{l_{ij}}S_{k_{ij}^{(l_{ij})}}(g_{ij}^{(l_{ij})},A_-^{(i)},A_+^{(j)}).
\ee
The variation of this action under the transformations \eqref{trans} is given by 
\be\label{var}
\d S_t=\sum_{i,j}\sum_{l_{ij}} {k_{ij}^{(l_{ij})}\ov 2\pi} \int d^2\s \
\Tr\big[ (A_+^{(i)} \del_- u_i - A_-^{(i)} \del_+ u_i) - (A_+^{(j)}  \del_- u_j - A_-^{(j)}  \del_+ u_j)\big]\ .
\ee
By exchanging $i\leftrightarrow j$ in the second parenthesis of \eqref{var} and by gathering identical terms we deduce that
\be\label{cond0}
\d S_t=0 \Longleftrightarrow \sum_{j,l_{ij}}k_{ij}^{(l_{ij})}=\sum_{j,l_{ji}}k_{ji}^{(l_{ji})},\,\,\, \forall \,i\, .
\ee
In the spirit of the discussion below \eqref{defj} this relation can be interpreted as level conservation at each site $i$.\footnote{ Here, by level conservation, we mean that the sum of the levels of the WZW models pointing to a specific vertex is equal to the sum of the levels pointing away from it. We will be using this jargon and hope that this abuse of language will not confuse the reader.} 
We have, thus, seen that gauge invariance of the action is equivalent to 
the requirement that the sum of levels that flows towards any site should be equal to the sum of levels that flows away from it. 
In passing, let us mention that the relation \eqref{cond0} implies that the action \eqref{full-act}  is not invariant only under the infinitesimal gauge transformations \eqref{trans} but also under the finite version of the gauge transformations.

At this point it would be useful to define the following quantities
\be\label{defk}
\tilde k_i=\sum_{j,l_{ij}}k_{ij}^{(l_{ij})}, \qquad \hat k_i=\sum_{j,l_{ji}}k_{ji}^{(l_{ji})}\, , \,\,\forall\,i\,.
\ee
One may now fix the gauge by choosing $\tilde g_i=\mathbb{1}$ for $i=1,2,\dots ,n$. Alternatively, one could have chosen to set to the unit element one or more of the group elements  $g_{ij}^{(l_{ij})}$ appearing in the WZW models leaving, as a result, some of the $\tilde g_i$ intact, that is leaving them as dynamical degrees of freedom. Notice that this is possible due to the fact that most of the WZW models are asymmetrically gauged. We have not checked explicitly but most probably, and up to global issues, this second choice should be related to the first one by a coordinate transformation, as it happens in the case of non-abelian T-duality. After the gauge fixing one ends up with the following action
\be\label{gfact}
\begin{split}
&S_{gf}=-{1\ov \pi}\int d^2\s\  A_{+\,a}^{(i)} \,\, (\l^{-1})_{ij}^{ab}\,\, A_{-\,b}^{(j)} +\\
&\sum_{i,j}\sum_{l_{ij}} \Bigg( S_{k_{ij}^{(l_{ij})}}(g_{ij}^{(l_{ij})})
+{k_{ij}^{(l_{ij})}\ov \pi} \int d^2\s \ \Tr \Big(A_-^{(i)} J_{+\,ij}^{(l_{ij})}   - A_+^{(j)} J_{-\,ij}^{(l_{ij})}+ A_-^{(i)} g_{ij}^{(l_{ij})}A_+^{(j)} \big(g_{ij}^{(l_{ij})}\big)^{-1}\Big) \Bigg),
\end{split}
\ee
where
\be\label{lambdadef}
(\l^{-1})_{ij}^{ab}={1 \ov 2}(\tilde k_i+\hat k_i)\,\d_{ij}\,\d ^{ab}+E_{ij}^{ab}=\d_{ij}\,(\l_{i}^{-1})^{ab} ,\qquad \,\,\,{\rm since}\,\,\,
E_{ij}^{ab}=\d_{ij} E_i^{ab}.
\ee

\begin{figure}[h]
\centering
\includegraphics[scale=1.3]{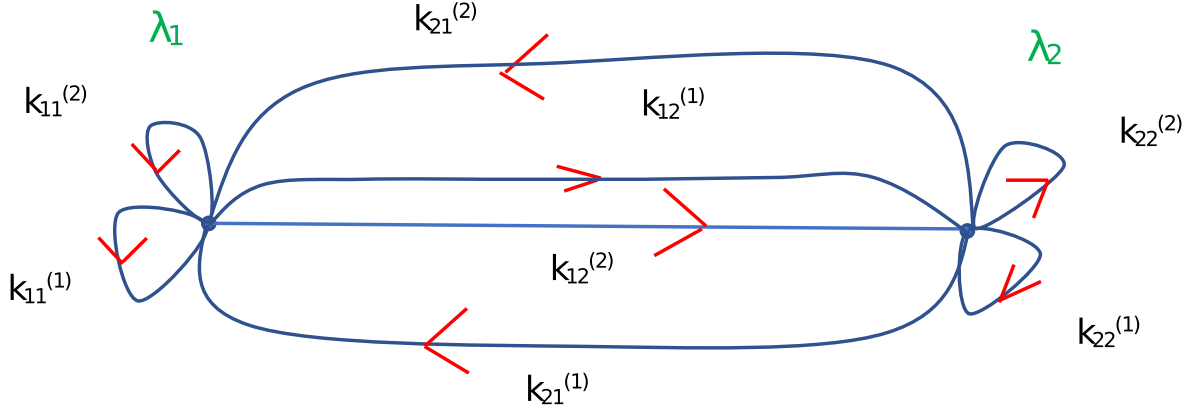}
\caption{{Diagram of an integrable web connecting any two of the fundamental  theories through four asymmetrically and four anomaly free gauged WZW models. The  two fundamental  theories are sitting at the two vertices. There are four different kinds of vertices corresponding to the four fundamental  theories, the PCM, the YB model, the isotropic $\s$-model on the symmetric space $G/H$ and the YB model on the symmetric space $G/H$.  The directed lines (blue lines with red arrows) connecting the vertices are associated with WZW models at arbitrary levels subject to the condition that  level conservation at each vertex is imposed. The diagram corresponds to an integrable theory with action of the form \eqref{gfact} and \eqref{sigma-fin}.}}
\label{fig:generaltriangle}
\end{figure}

We are now in position to present a diagrammatic representation of the action \eqref{gfact}. Namely,\\
\textbullet \,\,\, With every action functional of the form \eqref{gfact} we associate a certain diagram (see, for example, figure 1, figure 2 and figure 3 for the  integrable webs  connecting
 two, three  and four fundamental integrable theories, respectively).\\
\textbullet\,\,\, To each vertex of a diagram we assign the matrix $\l_i^{-1}$ of one of the fundamental integrable theories. These vertices represent the first line of \eqref{gfact}.
The number of vertices $n=n_1+n_2+n_3+ n_4$ is equal to the number of the fundamental integrable theories, 
that is $n_1$ different copies of the PCM, $n_2$ different copies of the YB model \footnote{Each of the YB models can have different parameters and different $R$ matrices obeying the modified YB equation.}, both based on a group $G$,  $n_3$ different copies of the symmetric coset space $G/H$ and $n_4$ different copies of the YB model on the symmetric space $G/H$. \\
\textbullet\,\,\, To each  line with orientation
 connecting two vertices $i$ and $j$ we assign one of the asymmetrically gauged WZW models in the second line of \eqref{gfact}. The directed line  is characterised by an integer number equal  to the level $k_{ij}^{(l_{ij})}$ of the WZW model with the flow of the level being from $i$ to $j$. \footnote{Notice that the flow of levels to the opposite direction from vertex $j$ to vertex $i$ is related to the WZW with group element  $g_{ji}^{(l_{ji})}$ at level $k_{ji}^{(l_{ji})}$.} There can be more than one 
lines
with the same direction connecting the vertex $i$ to the vertex $j$. The superscript $l_{ij}$ counts how  many different lines connecting $i$ and $j$ and having direction from $i$ towards $j$ the diagram has.\\
\textbullet\,\,\,  One may also have tadpole-like 
lines connecting the vertex $i$ to itself (see figures 1 and 2). In this case the corresponding WZW model is gauged in the usual anomaly free way. \\
\textbullet\,\,\,  Finally, as mentioned above, gauge invariance of the action, before fixing the gauge of course, is  equivalent to  level conservation at each vertex. This fact imposes $n-1$ constraints on the  levels circulating 
in the diagram, namely that $\tilde k_i=\hat k_i,\,\, \forall i$. Notice that if one imposes  level   conservation to $n-1$ vertices then  level  conservation of the remaining vertex is automatically satisfied.

\begin{figure}[h]
\centering
\includegraphics[scale=1.3]{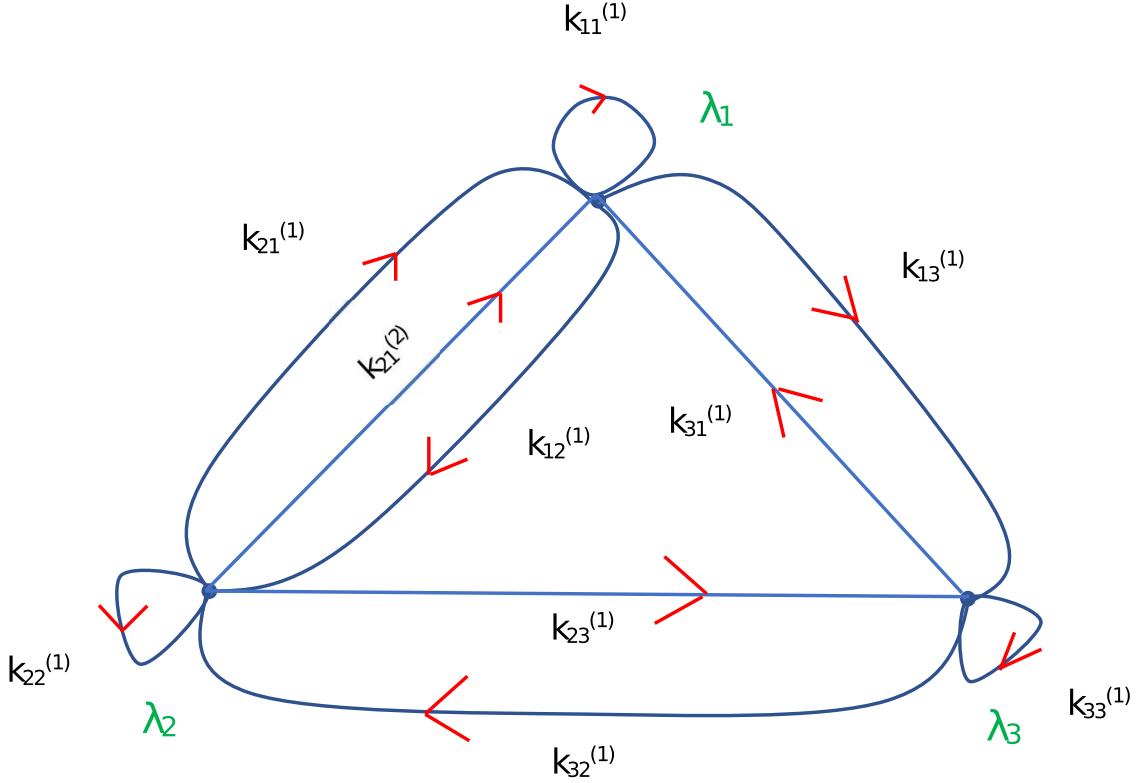}
\caption{{An example of an integrable web coupling any three of the fundamental  theories through seven asymmetrically and three vectorially gauged WZW models. The  three fundamental  theories are sitting at the three vertices. 
The  lines with orientation (blue lines with red arrows) connecting the vertices are associated with WZW models at arbitrary levels subject to the condition that  level conservation at each vertex is imposed. The diagram corresponds to an integrable theory.}}
\label{fig:generaltriangle}
\end{figure}

Let us now comment on figures 1, 2 and 3. The diagram of figure 1 has two vertices, i.e. $n=2$, and as a result represents  one of the ways to connect two of the fundamental theories. This connection is achieved through eight gauged WZW 
models four of which are asymmetrically gauged. Two of the latter have direction from the vertex/model 1 to the vertex/model 2 while the other two from the vertex/model 2 to the vertex/model 1. The four tadpole-like parts of the diagram correspond to four anomaly free vectorially gauged WZW models   connecting the two vertices to themselves.
Figure 2 depicts   an integrable theory consisting of 
 of three fundamental theories  $n=3$, seven asymmetrically and three vectorially gauged WZW models. Finally, figure 3 is an example of an integrable web that consists of four, i.e. $n=4$, of the fundamental
integrable theories where each of the vertices is connected to all others. This diagram has a total of sixteen lines with orientation/WZW models.

In order to obtain the $\s$-model action one should integrate out the gauge fields $A_\pm^{(i)}$ from \eqref{gfact}. To this end we evaluate
\be\label{eomAp}
{\d S_{gf}\ov \d A_+^{(i)}}=0\Longrightarrow    A_-^{(i)}  =-
\Big({1\ov \l^{-1}-{\mathcal D}^T}\Big)_{ij} \,  \mathbb J_-^{(j)}, \qquad                 \mathbb J_-^{(j)}=\sum_{n,l_{nj}} k_{nj}^{(l_{nj})} J_-(g_{nj}^{(l_{nj})} ) ,
\ee
and
\be\label{eomAm}
{\d S_{gf}\ov \d A_-^{(i)}}=0\Longrightarrow    A_+^{(i)}  =
\Big({1\ov \l^{-T}-{\mathcal D}}\Big)_{ij} \,  \mathbb J_+^{(j)}, \qquad                 \mathbb J_+^{(j)}=\sum_{ n,l_{jn}} k_{jn}^{(l_{jn})} J_+(g_{jn}^{(l_{jn})} ) ,
\ee
where we have also defined the matrices
\be\label{DDT}
{\mathcal D}_{ij}=\sum_{l_{ij}}  k_{ij}^{(l_{ij})}  D(g_{ij}^{(l_{ij})}),\qquad \mathcal D^T_{ij}=\sum_{l_{ji}}  k_{ji}^{(l_{ji})} D^T(g_{ji}^{(l_{ji})}),\qquad D^{ab}(g)=\tr{(t^a g \,t^bg^{-1}) }.
\ee
Thus, every entry of the matrix ${\cal D}_{ij}$ is the sum of the  $D^{ab}$ matrices of the group elements which connect the corresponding vertices/models weighted appropriately by their WZW levels.
At this point we should stress that the transpositions in \eqref{eomAp},  \eqref{eomAm} and  \eqref{DDT} apply only to the suppressed group indices $a,b=1,\ldots,dim(G) $.
Therefore, the entries of the matrices $\l^{-T}-{\cal D}$ and $\l^{-1}-{\cal D}^T$
are matrices themselves with their indices taking values in the group $G$.
Consequently, their inversion is to be understood as an inversion in the space of the fundamental integrable models keeping in mind that their entries are non-commutative objects.
One can now substitute the gauge fields \eqref{eomAp} and \eqref{eomAm} in \eqref{gfact} to get the $\s$-model
\be\label{sigma-fin}
\begin{split}
S_{\s-mod.}=-{1\ov \pi}\int d^2\s\   \mathbb J_+^{(i)} \,\, \,\, \Big({1\ov \l^{-1}-{\mathcal D}^T}\Big)_{ij} \,  \mathbb J_-^{(j)} +\sum_{i,j}\sum_{l_{ij}}  S_{k_{ij}^{(l_{ij})}}(g_{ij}^{(l_{ij})}).
\end{split}
\ee

As an example the matrix $\l^{-1}-{\mathcal D}^T$ corresponding to the diagram of figure  2 takes the form
\be\label{example}
\begin{split}
\l^{-1}-{\mathcal D}^T= 
\left(       \begin{array}{ccc}
            \l^{-1}_{1} -k_{11}^{(1)}D_{11}^T& -k_{21}^{(1)}D^{(1)\,T}_{21}-k_{21}^{(2)}D^{(2)\,T}_{21}&-k_{31}^{(1)}D_{31}^T  \\
            -k_{12}^{(1)}D_{12}^T &  \l^{-1}_{2} -k_{22}^{(1)}D_{22}^T   &  -k_{32}^{(1)}D_{32}^T \\
              -k_{13}^{(1)}D_{13}^T   &  -k_{23}^{(1)}D_{23}^T &    \l^{-1}_{3} -k_{33}^{(1)}D_{33}^T\\
           \end{array}
         \right)\  .
\end{split}
\ee

An important comment is in order. Note that although \eqref{sigma-fin} is similar in form to equation (2.13) of \cite{Georgiou:2018gpe}, the models of the present work certainly do not  belong to the subclass of the {\it integrable}  sector of the models presented in  \cite{Georgiou:2018gpe} and most probably they do not belong at all to the general class of the models constructed in \cite{Georgiou:2018gpe}.
A first hint comes from inspecting of the matrix $\cal D$ which in our case may generically have non-zero entries everywhere in contradistinction to  \cite{Georgiou:2018gpe} where $\cal D$ has entries only along the diagonal. 
The reason behind this difference is that in the models of \cite{Georgiou:2018gpe}, as well as in those of \cite{Bassi:2019aaf} the number of the WZW terms is equal to that of the integrable theories one couples while in our case the number of the WZW models 
is strictly greater or equal to that of the integrable theories we couple. 

\subsection{Reading the $\s$-model from its diagram}

We are now in position to reverse the argument. Given any diagram one can immediately write down the corresponding integrable $\s$-model action. The steps are as follows.\\
\textbullet\,\,\, Draw a diagram with any number of vertices and to each vertex assign one of the four fundamental integrable theories (see, for example, figure 3).\\
\textbullet\,\,\, Connect the vertices with any number of directed lines
you wish in such a way that  level conservation  at each vertex holds.\\
\textbullet\,\,\, For each directed line write a WZW model at the level dictated by the level   of the  line  ($2^{nd}$ term in \eqref{copy-sigma-fin}).\\
\textbullet\,\,\, For each vertex write the incoming and outgoing currents $\mathbb J_-^{(i)}$ and $\mathbb J_+^{(i)}$, respectively.\footnote{The expressions for $\mathbb J_-^{(i)}$and $\mathbb J_+^{(i)}$ can be found in \eqref{eomAp} and \eqref{eomAm}, respectively.}\\
\textbullet\,\,\, Finally, couple these currents through the matrix $\Big(\l^{-1}-{\mathcal D}^T\Big)^{-1} $, where $\cal D$ is defined in \eqref{DDT}, to get the special case of \eqref{sigma-fin} that corresponds to the diagram at hand.\\
\begin{figure}[h]
\centering
\includegraphics[scale=1.3]{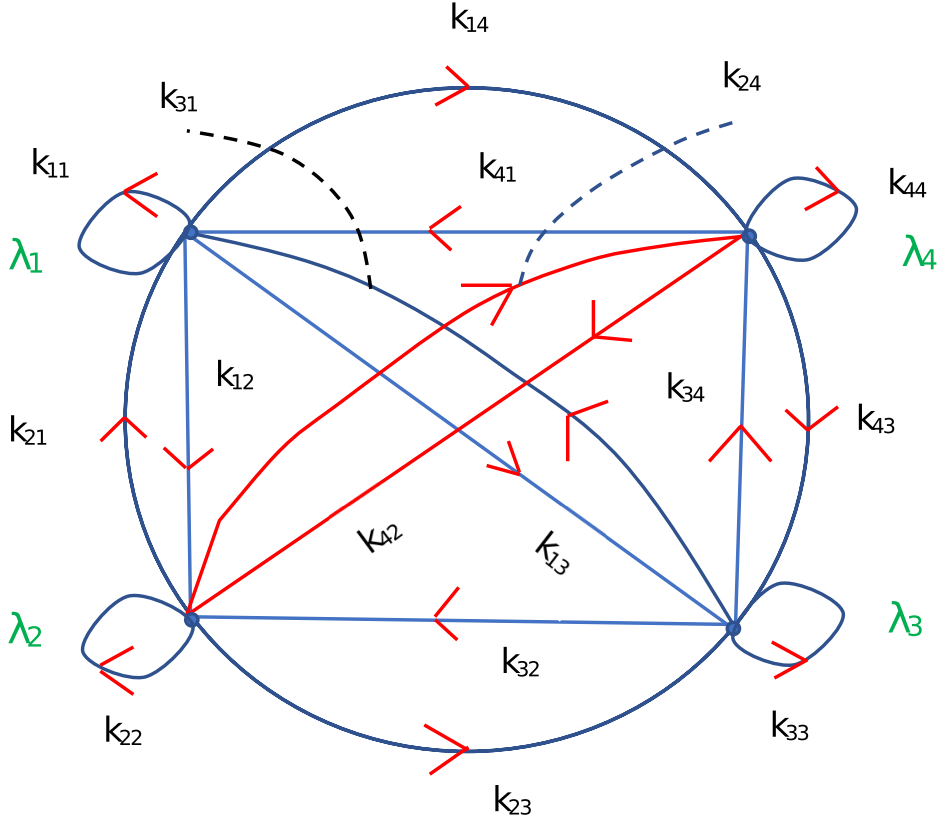}
\caption{{An example of an integrable web representing   an integrable theory built from four of the fundamental  theories, twelve asymmetrically and four vectorially gauged WZW models. The  four fundamental  theories are sitting at the four vertices. 
The lines with orientation (blue lines with red arrows) connecting the vertices are associated with WZW models at arbitrary levels subject to the condition that level conservation at each vertex is imposed. The intersections of the red with the blue  lines do not represent vertices. The red lines have no special meaning. They are drawn red so that we can stress that their intersections with the blue lines do not designate vertices. The diagram corresponds to an integrable theory.}}
\label{fig:generaltriangle}
\end{figure}
For the convenience of the reader we copy from the previous section the final form of the $\s$-model action
\be\label{copy-sigma-fin}
\begin{split}
S_{\s-mod.}=-{1\ov \pi}\int d^2\s\   \mathbb J_+^{(i)} \,\, \,\, \Big({1\ov \l^{-1}-{\mathcal D}^T}\Big)_{ij} \,  \mathbb J_-^{(j)} +\sum_{i,j}\sum_{l_{ij}}  S_{k_{ij}^{(l_{ij})}}(g_{ij}^{(l_{ij})}).
\end{split}
\ee

We should, of course, mention that the inverse of the matrix $\l^{-1}-{\mathcal D}^T$ has to be evaluated in a case by case basis. 
Finally, let us mention that for small values of the entries of the matrix $\big(\l^{-1}\big)^{-1}_{ij}$  the action becomes
\be\label{small-lam}
S_{\s-mod.}=-{1\ov \pi}\int d^2\s\   \mathbb J_+^{(i)} \,\, \,\, \big(\l^{-1}\big)^{-1}_{ij}\,  \mathbb J_-^{(j)} +\sum_{i,j}\sum_{l_{ij}}  S_{k_{ij}^{(l_{ij})}}(g_{ij}^{(l_{ij})})+{\cal O}(\l^2).
\ee
Notice that in our conventions $-i \, \mathbb J_-^{(i)}$ and $-i \, \mathbb J_+^{(i)}$ generate two Kac-Moody currents at levels $\hat k_i$ and $\tilde k_i$ respectively. As discussed above, these levels are equal due to level conservation. 
Finally, the same arguments with those presented in Appendix A
apply to the linearised action \eqref{small-lam}. When the first term in \eqref{small-lam} is written in a form involving an $N\times N$ matrix, where $N$ is the number of the WZW models, the corresponding $\big(\L^{-1}\big)^{-1}_{ij}$ where now $i,j=1,\dots,N$ is a non-invertible matrix. This means that there is no matrix $\L^{-1}$ with regular entries such that our model can be straightforwardly obtained from the models of \cite{Georgiou:2018gpe}. The same holds for the integrable models in section 4.1 of \cite{Bassi:2019aaf} which have the same structure as those in \cite{Georgiou:2018gpe} but with a more general $\L$ matrix. Last, but not least, we would like to stress that in the construction of the present paper some of the fundamental theories that serve as building blocks of the  final integrable theory are  the isotropic $\s$-models on the symmetric space $G/H$ or  the YB models on the symmetric space $G/H$. This was not the case neither in the construction of \cite{Bassi:2019aaf} nor in that of \cite{Georgiou:2018gpe}.\footnote{It might be possible that the models of the present work could be obtained as {\it
special decoupling limits} of those in the \cite{Bassi:2019aaf} (see also \cite{Delduc:2019bcl}) but only in the case where {\it all} the fundamental theories we couple are of the PCM-type. Finally, it would be interesting to see if our model can fit in the framework of \cite{Costello:2019tri}.}

\section{Proof of integrability}
In this section, we prove that the theories constructed in the previous section are integrable in the case where the coupling $\l^{-1}_{ij}=\d_{ij} \l_i^{-1}$. However in most of the manipulations and in anticipation of the results of the following sections 
we will treat $\l^{-1}_{ij}$ as being a general matrix.

The proof of integrability requires two steps. In section 3.1 we show that a subset of the equations, namely $n$  of them, can be recast as zero curvature conditions of certain Lax pairs which we explicitly find. One can then use construct the monodromy matrices whose traces are conserved for all values of the spectral parameter. By expanding the monodromy matrices one can obtain $n$ infinite towers of conserved charges. In general, these conserved charges are not a-priori in involution. One case where these charges are in involution is when the spatial part of the Lax connections assumes the $r/s$ Maillet form.  At the end of section 3.1 we show that the charges obtained from different Lax pairs are in involution. However, we have not checked if the charges obtained from the same Lax connection are in involution among themselves. We believe that this will be the case.

Notice that the total number $n$ of Lax pairs given in \eqref{Lax} is generically  smaller than  $N$ which is the number of the degrees of freedom of the theory.\footnote{Strictly speaking the number of degrees of freedom is $N \,dim(G)$. For the sake of brevity we will refrain from referring to the $dim(G)$ factor.} The latter can be taken to be  the group elements $g_{ij}^{(l_{ij})}$. This mismatch happens because the number of vertices is generically smaller than the number of directed lines (see any of the diagrams). 
As a result, the fact that it is only a certain combination \eqref{inter-eom} of the full set of eoms \eqref{eomWZW} that are equivalent to the Lax equations \eqref{Lax} indicates that the conserved charges obtained from these Lax pairs are not enough to ensure integrability.

In Section 3.2, we show that each of the remaining $N-n$ equations of motion can be brought to the form of a covariantly free quantity. This straightforwardly implies the existence of $N-n$ infinite towers of conserved charges which supplement those obtained from the Lax connections. As a result one has as many infinite towers of conserved charges as the number of degrees of freedom of the theory. This concludes our proof of integrability.

Finally, let us make the following comment. The situation encountered in the case of our models is slightly peculiar and certainly interesting. As mentioned above,
it is not the full set of equations of motion that can be rewritten as Lax pairs satisfying 

flatness conditions. Instead, part of the equations of motion are equivalent to Lax equations while the remaining ones take the form of covariantly free quantities. The latter equations lead  to the direct construction of infinite towers of conserved charges. It is true that the covariantly free equations of motion can not be put in the Lax form.
However, the primary and most fundamental definition of an integrable system is that one can find conserved charges  \footnote{ Which actually should be in involution (see also the discussion at the end of section 3.2.)},  the number of which
should be  equal to the degrees of freedom of the theory under consideration. In this section, we show that our models have this property.
The situation encountered here is similar to that of the WZW model.\footnote{We thank the referee for pointing out this similarity.} Actually, we think that the models presented in this work are canonically equivalent to the sum of $n$ single $\lambda$-deformed models plus $N-n$ gauged WZW models \cite{Georgiou}.

\subsection{Lax connections for a subset of equations of motion }
We start with the equations of motion for $A_\pm^{(i)}$. These can be easily brought to the form
\be\label{Am}
\sum_{i, l_{ij}}k_{ij}^{(l_{ij})}\big(g_{ij}^{(l_{ij})}\big)^{-1}D_-g_{ij}^{(l_{ij})}=-(\l^{-1}_{jn}-\hat k_n \d_{jn})A_-^{(n)}
\ee
and 
\be\label{Ap}
\sum_{j, l_{ij}}k_{ij}^{(l_{ij})}D_+g_{ij}^{(l_{ij})}\big(g_{ij}^{(l_{ij})}\big)^{-1}=(\l^{-T}_{ni}-\tilde k_n \d_{in})A_+^{(n)},
\ee
where $\tilde k_n$ and $\hat k_n$ are defined in \eqref{defk} and the transposition in 
\eqref{Ap} refers only to the suppressed group indices. Notice also that although $\tilde k_n=\hat k_n$ due to  level conservation we have not imposed this 
condition in  \eqref{Am} and  \eqref{Ap} yet. Finally, the covariant derivatives on the WZW group elements read $D_\pm g_{ij}^{(l_{ij})}= \partial_\pm g_{ij}^{(l_{ij})}-A_\pm^{(i)}g_{ij}^{(l_{ij})}+g_{ij}^{(l_{ij})}A_\pm^{(j)}$.

In addition, we will  need the equations of motion for the group elements of the  WZW models. These turn out to be
\be\label{eomWZW}
{\d S_{gf}\ov \d g_{ij}^{(l_{ij})}}=0\Longrightarrow D_-\Big(D_+g_{ij}^{(l_{ij})}\big(g_{ij}^{(l_{ij})}\big)^{-1}\Big)=F_{+-}^{A^{(i)}}\Longleftrightarrow  D_+\Big(\big(g_{ij}^{(l_{ij})}\big)^{-1}D_-g_{ij}^{(l_{ij})}\Big)=F_{+-}^{A^{(j)}},
\ee
where the field strenghts are defined as usual, $F_{+-}^{A^{(i)}}=\partial_+ A_-^{(i)}- \partial_- A_+^{(i)}-[A_+^{(i)},A_-^{(i)}]$ and where the left covariant derivative in the second and third equation of \eqref{eomWZW} are acting to its arguments according to their transformation properties, namely $D_-\cdot=  \partial_-\cdot -[A_-^{(i)},\cdot]$ and $D_+\cdot=  \partial_+\cdot -[A_+^{(j)},\cdot]$ respectively.
Multiplying the second and third equation in \eqref{eomWZW} by $k_{ij}^{(l_{ij})}$  and summing over $j, l_{ij}$ and $i, l_{ij}$ respectively we arrive at 
\be\label{inter-eom}
\begin{split}
&\sum_{j, l_{ij}}k_{ij}^{(l_{ij})}D_-\Big( D_+g_{ij}^{(l_{ij})}\big(g_{ij}^{(l_{ij})}\big)^{-1}\Big)=\tilde k_i F_{+-}^{A^{(i)}}\, ,\\
&\sum_{i, l_{ij}}k_{ij}^{(l_{ij})}D_+\Big( \big(g_{ij}^{(l_{ij})}\big)^{-1}D_-g_{ij}^{(l_{ij})}\Big)=\hat k_jF_{+-}^{A^{(j)}}.
\end{split}
\ee
Substituting now \eqref{Am} and \eqref{Ap} in \eqref{inter-eom} we get after some algebra the equations of motion of the system expressed solely in terms of the gauge fields. These read
\be\label{semi-eom}
\begin{split}
&\tilde k_i\,\del_+A_-^{(i)}- \l_{ni}^{-T}\del_-A_+^{(n)}=[\l_{ni}^{-T}A_+^{(n)},A_-^{(i)} ]\, ,\\
&\l_{in}^{-1}\del_+A_-^{(n)}- \hat k_i\,\del_-A_+^{(i)}=[A_+^{(i)},\l_{in}^{-1}A_-^{(n)} ]\,.
\end{split}
\ee
In the case where $\l_{ij}^{-1}=\d_{ij}\l_{i}^{-1}$ and the  levels   at each vertex are conserved, i.e. $\tilde k_i=\hat k_i$, which is precisely the case we consider in this section, the equations in 
\eqref{semi-eom} decouple in the space of models and become
\be\label{semi-eom1}
\begin{split}
&\del_+A_-^{(i)}-\hat \l_{i}^{-T}\del_-A_+^{(i)}=[\hat\l_{i}^{-T}A_+^{(i)},A_-^{(i)} ]\, ,\\
&{\hat \l}_{i}^{-1}\del_+A_-^{(i)}- \del_-A_+^{(i)}=[A_+^{(i)},\hat \l_{i}^{-1}A_-^{(n)} ]\, ,
\end{split}
\ee
where $\hat \l_{i}^{-1}=\mathbb 1+{E_i\ov k_i}=\frac{\l_i}{k_i}$. In the last relation we have used  level   conservation to define $k_i=\hat k_i=\tilde k_i $.
Thus we see that the equations of motion of our theory reduce to $n=n_1+n_2+n_3+n_4$ decoupled sets of equations. Each of these sets correspond to the equations of motion of a single $\l$-deformed model with $n_1$ of the  $\hat \l_{i}^{-1}$s being the isotropic matrices of PCMs, $n_2$ of the  $\hat \l_{i}^{-1}$s being the  matrices of YB models, $n_3$ of the  $\hat \l_{i}^{-1}$s being the isotropic matrices of a symmetric coset space $G/H$ and  $n_4$ of the  $\hat \l_{i}^{-1}$s being the matrices of the YB model on the symmetric space $G/H$.

Notice that despite the decoupling of the equations of motion when these are expressed in terms of the gauge fields, the $\s$-model action assumes a non-trivial form in which the group elements and the deformation matrices $\hat \l_{i}^{-1}$ are coupled in a very complicated way the details of which depend on the topology of the corresponding diagram. More precisely, to the same set of $n=n_1+n_2+n_3+n_4$ fundamental integrable models and WZW models at certain levels 
correspond many different coupled $\s$-models since there are many different ways/diagrams to connect the vertices (fundamental models) with the  directed lines
(asymmetrically gauged WZW models). Furthermore, the Hamiltonian density 
of our models can not be written solely in terms of the gauge fields $A_\pm^{(i)}$ and the couplings ${\hat \l}_{i}^{-1}$ as it was possible in the case of doubly and cyclic $\l$-deformed models which were shown to be canonically equivalent to the sum of two or more single $\l$-deformed models \cite{Georgiou:2016urf,Georgiou:2017oly}. 
This essential  difference can be traced to the fact that in our models the group degrees of freedom $g_{ij}^{(l_{ij})}$ are generically strictly greater than the number of the fundamental theories we couple and thus greater than the number of the gauge fields of the theory (see for example \eqref{full-act} or \eqref{gfact}).

 Consider now a diagram in which all possible kinds of lines and vertices are present, that is when each vertex is  connected to itself and to all other vertices in both directions.
The number of independent parameters that the  model corresponding to such a diagram possess is at least $n^2+1+n_2+n_4$.
This number comes out as follows. Since each of the vertices can be connected to all other vertices including itself  the most generic diagram \footnote{ As mentioned above, as most generic we characterise a diagram in which all possible kinds of lines and vertices are present.} has at least $n^2$ parameters which are the levels $k_{ij}^{(l_{ij})}$.  Level conservation imposes $n-1$ constraints 
on the levels (see last bullet point below \eqref{lambdadef}). Lastly, one has $n+n_2+n_4$ continuous parameters in the definitions of the $\hat \l_{i}^{-1}$ matrices. Putting everything together we get that our models depend on at least $n^2+1+n_2+n_4$ parameters.

Equations \eqref{semi-eom1} imply the existence of $n$ independent Lax pairs satisfying
\be\label{Lax}
\del_+{\cal L}^{(i)}_- -\del_-{\cal L}^{(i)}_+-[{\cal L}^{(i)} _+,{\cal L}^{(i)} _-]=0.
\ee
These are given by 
\be\label{iso}
{\cal L}^{(i)}_\pm=
{2 \ov 1+\hat \l_{i}}{z_i \ov 1\mp z_i}A_\pm^{(i)},\,\,\,\qq1\leq i\leq n_1
\ee
in the case where $\hat \l_{i}$ is the coupling obtained from the isotropic matrix of a PCM, whereas in the case where 
$\hat \l_{i}=\mathbb 1+{1\ov k_i\,t_i}(\mathbb 1-\eta_i {\cal R}_i)^{-1}$ and the coupling is obtained from a YB model the Lax pair is given by \cite{Sfetsos:2015nya}
\be\label{YB}
\begin{split}
&{\cal L}^{(i)}_\pm=\big((\a^{(i)}_1+\a^{(i)}_2{z_i \ov z_i\mp 1 })\mathbb 1\pm\eta_i {\cal R}_i\big)\big(\mathbb 1\pm\eta_i {\cal R}_i\big)^{-1}A_\pm^{(i)},\,\,\,\qq n_1<i\leq n_1+n_2\\
&\a^{(i)}_1=\a_i-\sqrt{\a^2-c^2 \eta_i^2},\qq \a^{(i)}_2=2\sqrt{\a_i^2-c^2 \eta_i^2},\qq \a_i={1+c^2 \eta_i^2 \rho_i\ov 1+\rho_i},\qq \rho_i={k_i t_i\ov 1+k_i t_i}.
\end{split}
\ee
In \eqref{YB} $c^2=0,\pm1$ and the skew symmetric  matrix ${\cal R}_i$ satisfies the modified Yang-Baxter equation $[{\cal R}_iA,{\cal R}_i B]-
{\cal R}_i([{\cal R}_iA,B]+[A,{\cal R}_iB])=-c^2[A,B],\,\, \forall A,B \in {\cal L}(G)$.
The third possibility is when the fundamental integrable model sitting at a vertex is the isotropic $\s$-model on a symmetric space of the coset form $G/H$.
In this case the equation of motion \eqref{semi-eom1} become
\be\label{coset-eom}
\begin{split}
&\del_\pm A_\mp^{(i)g/h}=-[A_\mp^{(i)g/h},A_\pm^{(i)h}], \,\, \del_+ A_-^{(i)h}-\del_- A_+^{(i)h}-[A_+^{(i)h},A_-^{(i)h}]={1\ov  \hat\l_i}[A_+^{(i)g/h},A_-^{(i)g/h}],\\
&A_\pm^{(i)}=A_\pm^{(i)h}+A_\pm^{(i)h},\,\, A_\pm^{(i)h} \in {\cal L}(H),\, A_\pm^{(i)g/h} \in {\cal L}(G/H), \,\,\,\,\,\,\,\,\,n_1+n_2<i \leq n_1+n_2+n_3.
\end{split}
\ee
These equations imply the existence of a Lax connection of the following form \cite{Hollowood:2014rla}
\be
{\cal L}_\pm^{(i)}=A_\pm^{(i)h}+{z_i^{\pm1} \ov \sqrt{ \hat\l_i}} A_\pm^{(i)g/h},\,\,\,\qq n_1+n_2<i \leq n_1+n_2+n_3
\ee
where $z_i$ is, as usual, the spectral parameter.\\
The fourth and last possibility is when the fundamental integrable model sitting at a vertex is a YB model based on the symmetric space of the coset form $G/H$.
In this case the equation of motion \eqref{semi-eom1} imply the existence of a Lax connection of the form \cite{Sfetsos:2015nya}
\be
{\cal L}_\pm^{(i)}=A_\pm^{(i)h}+z_i^{\pm1}({1 \ov \sqrt{\rho_i}} +\eta_i \, \rho_i^{\pm{1\ov 2}} {\cal R}_i)\big(\mathbb 1\pm\eta_i {\cal R}_i\big)^{-1}A_\pm^{(i)g/h},\,\,\,\qq n_1+n_2+n_3<i \leq n
\ee
given that the projection of the ${\cal R}_i$-bracket in the sub-algebra $h$ vanishes, namely that 
\be \label{cond}
([{\cal R}_i X,Y]+[X,{\cal R}_i Y])|_h=0, \qq X,Y\in g/h\, .
\ee
We close this section with an important comment. The infinite tower of conserved charges obtained from any of the above $n=n_1+n_2+n_3+n_4$ Lax pairs are in involution with those obtained from all the remaining $n-1$ Lax pairs. To see this one can define the following dressed currents that obey commuting Kac-Moody algebras \cite{Bowcock,Georgiou:2016urf,Georgiou:2017oly}
\be\label{drcurr}
\begin{split}
&{\cal J}_{-\,ji}^{(l_{ji})}=-\big(g_{ji}^{(l_{ji})}\big)^{-1}D_-g_{ji}^{(l_{ji})}+A_-^{(i)}-A_+^{(i)}\\
&{\cal J}_{+\,ij}^{(l_{ij})}=D_+g_{ij}^{(l_{ij})}\big(g_{ij}^{(l_{ij})}\big)^{-1}+A_+^{(i)}-A_-^{(i)}.
\end{split}
\ee
Multiplying the first equation by $k_{ji}^{(l_{ji})}$ and summing over $j$ and $l_{ji}$ and the second equation by $k_{ij}^{(l_{ij})}$ and summing again over $j$ and $l_{ij}$ one obtains, after substituting in the constraints \eqref{Am} and \eqref{Ap}, the following relations
\be\label{drcurr1}
\begin{split}
&\sum_{j, l_{ji}}k_{ji}^{(l_{ji})}{\cal J}_{-\,ji}^{(l_{ji})}=\big(\l_i^{-1}A_-^{(i)}- \hat k_iA_+^{(i)}\big)\\
&\sum_{j, l_{ij}}k_{ij}^{(l_{ij})}{\cal J}_{+\,ij}^{(l_{ij})}=\big(\l_i^{-T}A_+^{(i)}- \tilde k_i A_-^{(i)}\big).
\end{split}
\ee
This set of equations can be now solved for the gauge fields $A_\pm^{(i)}$  in terms of $\sum_{j, l_{ji}}k_{ji}^{(l_{ji})}{\cal J}_{-\,ji}^{(l_{ji})}$ and $\sum_{j, l_{ij}}k_{ij}^{(l_{ij})}{\cal J}_{+\,ij}^{(l_{ij})}$. Given that the Poisson brackets $\{{\cal J}^{(l_{ij})}_{+\,ij},{\cal J}^{(l_{\hat ik})}_{+\,\hat i k}\}_{PB}=0=\{{\cal J}^{(l_{ji})}_{-\,ji},{\cal J}^{(l_{k\hat i})}_{-\,k\hat i }\}_{PB}$ when $i \neq \hat i$ and that $\{{\cal J}^{(l_{..})}_{+\,. .},{\cal J}^{(l_{..})}_{-\,..}\}_{PB}=0$ we deduce that $\{A_\pm^{(i)},A_\pm^{(\hat i)}\}_{PB}=0\,\,\, {\rm for}\,\,\,i\neq \hat i$. As a result of the last equation we have that  $\{{\cal L}_\pm^{(i)},{\cal L}_\pm^{(\hat i)}\}_{PB}=0 \,\,\, {\rm for} \,\,\,i\neq \hat i$ which in turn implies that the conserved charges obtained from 
different Lax connections are in involution.

\subsection{Infinite towers of conserved charges from the remaining equations of motion}

In this section, we will argue that, besides the charges that can be obtained from the Lax pairs of the previous  section, there are $N-n$ additional  towers of infinite conserved charges.
The argument goes as follows.
Consider any of the $n$ vertices and focus on the directed lines/WZW models that are pointing away from  this vertex. Next choose one of the latter, say the one depending on $g_{ij}^{( l^{0}_{ij})}$ as a reference. Then the full set of eoms  $D_-\Big(D_+g_{ij}^{(l_{ij})}\big(g_{ij}^{(l_{ij})}\big)^{-1}\Big)=F_{+-}^{A^{(i)}}$ in \eqref{eomWZW} can be rewritten as
\be\label{eomWZW-1}
\begin{split}
&\sum_{j, l_{ij}}k_{ij}^{(l_{ij})}D_-\Big( D_+g_{ij}^{(l_{ij})}\big(g_{ij}^{(l_{ij})}\big)^{-1}\Big)=\tilde k_i F_{+-}^{A^{(i)}}\, ,\\
&D_-Y_{+\,\,ij}=0, \qquad {\rm where}\qquad
Y_{+\,\,ij}=D_+g_{ij}^{(l_{ij})}\big(g_{ij}^{(l_{ij})}\big)^{-1}-D_+g_{ij}^{(l^{0}_{ij})}\big(g_{ij}^{(l^{0}_{ij})}\big)^{-1}.
\end{split}
\ee
We see, thus, that the complete set of equations is equivalent to eq.  \eqref{inter-eom}, which is the one used for the construction of the Lax pairs \eqref{Lax}, plus a number of covariantly free currents given by the second relation in \eqref{eomWZW-1}.  Now each of the covariantly free currents generates an infinite tower of conserved charges given by
\be\label{charges}
Q_{ij}^{(k)}=\int_0^{2 \pi} \,d\s\, \,\tr(Y_{+\,\,ij})^k,\,\,\,k=1,2,3,\dots .
\ee
Indeed, 
\be\label{proof}
\begin{split}
&\del_\tau Q_{ij}^{(k)}= k \int_0^{2 \pi} \,d\s \,\, \tr\Big( (Y_{+\,\,ij})^{k-1} \del_\tau Y_{+\,\,ij}\Big)
=k \int_0^{2 \pi} \,d\s \,\, \tr\Big( (Y_{+\,\,ij})^{k-1} \del_\s Y_{+\,\,ij}\Big)+\\
&+2 k \int_0^{2 \pi} \,d\s \,\, \tr\Big( (Y_{+\,\,ij})^{k-1}[A_-^{(i)},Y_{+\,\,ij}]\Big)\, ,
\end{split}
\ee
where we have used the fact that $Y_{+\,\,ij}$ is covariantly conserved, that is ${1\ov 2}(\del_\tau-\del_\s )Y_{+\,\,ij}-[A_-^{(i)},Y_{+\,\,ij}]=0$. Then by rewriting $k \,\tr\Big( (Y_{+\,\,ij})^{k-1} \del_\s Y_{+\,\,ij}\Big)=\del_\s  \tr\Big( (Y_{+\,\,ij})^{k} \Big)$ and taking into account that $Y_{+\,\,ij}$ is a periodic function we deduce that \footnote{The last term of \eqref{proof} vanishes due to the cyclicity of the trace.}
\be\label{conserved}
\del_\tau Q_{ij}^{(k)}=0.
\ee
Notice that the total number of the towers of the conserved charges $Q_{ij}^{(k)}$ is equal to the number of the groups elements 
$g_{ij}^{(l_{ij})}$ minus the number of the vertices $n$ since, as can be seen from \eqref{eomWZW-1},  one equation per vertex can not be brought to the form of a covariantly free quantity. The conserved charges for this single equation are supplied by the Lax pairs of 
\eqref{Lax}.

We conclude that we have enough conserved charges to ensure integrability of the theories since one can construct as many infinite towers of charges as the number of the degrees of freedom of the theories which is, of course, equal to the number of the group elements $g_{ij}^{(l_{ij})}$. A last comment is in order. It is true that the local charges belonging to the same infinite tower
as defined in \eqref{charges} are not in involution.\footnote{Any two charges belonging to different towers are in involution because they depend on completely different group elements. }
However, one can employ a construction similar to the one in \cite{Evans:1999mj} to find certain combinations of the charges in \eqref{charges} which are in involution. As discussed in \cite{Evans:1999mj} the details of the construction depends on which precisely the group $G$ is.

\section{Coupling isotropic integrable theories}
In this section, we consider the special case where all the vertices are of the PCM-type, that is
 when the coupling matrices 
$(\l^{-1})^{ab}_{ij}$ are diagonal and isotropic in the group space, namely $(\l^{-1})^{ab}_{ij}=(\l^{-1})_{ij}\d^{ab}$. 
In section 4.1, we will consider the integrable case where the deformation matrix is non-diagonal in the space of theories with momentum conservation 
imposed at each vertex. In the next section 4.2, we will focus on the integrable case where the deformation matrix is diagonal in both the group space and the 
space of models, that is when $(\l^{-1})^{ab}_{ij}=\l_i^{-1}\d_{ij}\d^{ab}$, but  level  conservation at the vertices is not imposed.

\subsection{Non-diagonal in the space of models deformation matrix}
In this case after we make the following redefinitions 
\be\label{redefA}
\tilde A_-^{(i)}=\sqrt{\tilde k_i} A_-^{(i)},\qq \tilde A_+^{(i)}=\sqrt{\hat k_i} A_+^{(i)},\qq \tilde \l^{-1} _{ij}={1\ov \sqrt{\tilde k_i \hat k_i}}\l^{-1} _{ij}
\ee
the general equations of motion \eqref{semi-eom} become
\be\label{semi-final}
\begin{split}
&\del_+\tilde A_-^{(i)}- \tilde \l_{ji}^{-T}\del_-\tilde A_+^{(j)}={1 \ov \sqrt{\tilde k_i}}[\tilde \l_{ji}^{-T}\tilde A_+^{(j)},\tilde A_-^{(i)} ]\, ,\\
&\tilde \l_{ij}^{-1}\del_+\tilde A_-^{(j)}- \del_-\tilde A_+^{(i)}={1 \ov \sqrt{\hat k_i}}[\tilde A_+^{(i)},\tilde \l_{ij}^{-1}\tilde A_-^{(j)} ]\,.
\end{split}
\ee
If we now impose  level   conservation $k_i=\tilde k_i=\hat k_i$ we see
that in the case of isotropic 
$\l$ the equations of motion of our model take precisely the form of the equations of motion of the most general $\l$-deformed model constructed in \cite{Georgiou:2018gpe} (see eq. (2.9) of this work). This by no means that these two classes of theories are trivially identical since, as mentioned above, in our construction the number of WZW models is strictly greater or equal to the number of gauge fields while in the construction of \cite{Georgiou:2018gpe} these two number are precisely equal. Thus, the degrees of freedom of our models are generically greater than those 
of the models in \cite{Georgiou:2018gpe} for the same number of gauge fields. 

Given the form of the equations of motion \eqref{semi-final} we immediately deduce that when the matrix $\tilde \l^{-1}$ has the form \footnote{Notice that the $ \l^{-1}_{ij}$ in \eqref{lambda22} that follow are not the same with the $ \l^{-1}_{ij}$ of \eqref{redefA}. We have used the same letter so that we do not have proliferation of symbols.}
\be
\label{lambda22}
\begin{split}
\tilde \l^{-1}_{ij} =
\left(       \begin{array}{cccc}
            \l^{-1}_{11} & 0 &  \cdots  &0  \\
             \l^{-1}_{21} & 0 &\cdots  & 0\\
              \vdots  & \vdots  & \ddots & \vdots \\
              \l^{-1}_{(n-1)1}  & 0 & \cdots &  0\\
               0 & \l^{-1}_{n2} & \cdots  & \l^{-1}_{nn}\\
           \end{array}
         \right)\  .
\end{split}
\ee
the theory is integrable, exactly as it happened in \cite{Georgiou:2018gpe}, with the Lax pairs given by 
\be
\begin{split}
\label{Lax1}
\cL^{(1)}_+=\sum_{i=1}^{n-1}c_+^{(i)}(z) \tilde  A_+^{(i)} ,\qquad
\cL^{(1)}_-=z \tilde A_-^{(1)} ,
\end{split}
\ee
where 
\be
\begin{split}\label{c-sol}
&c_+^{(i)}={\l^{-1}_{i1}(\l^{-1}_{i1}-\mu_{i1}) \ov(\l^{-1}_{i1} -z\sqrt{k_i})}\ {z \ov d+d_1} \, 
i=1,2,\dots ,n-1\ ,\ \  d_1=\sum_{j=1}^{n-1} {\l^{-2}_{j1}(\l^{-1}_{j1}-\mu_{j1} )\ov \l^{-1}_{j1} -z\sqrt{k_j} },\,\,\,\mu_{i1}=\sqrt{{k_i \ov k_1}}
\end{split}
\ee
and 
\be
\begin{split}
\label{Lax2}
\cL^{(2)}_-=\sum_{i=2}^{n}c_-^{(i)} \tilde  A_-^{(i)} ,\qquad
\cL^{(2)}_+=z \tilde A_+^{(n)} ,
\end{split}
\ee
where
\be
\begin{split}\label{c-sol2}
&c_-^{(i)}={\l^{-1}_{ni}(\l^{-1}_{ni}-\mu_{ni}) \ov(\l^{-1}_{ni} -z\sqrt{k_i})}\ {z \ov \hat d + \hat d_1}\ ,\quad
i=2,\dots,n\ ,\quad  \hat d_1=\sum_{j=2}^n {\l^{-2}_{nj}(\l^{-1}_{nj}-\mu_{nj} )\ov \l^{-1}_{nj} -z\sqrt{k_j} },\,\,\,\mu_{ni}=\sqrt{{k_i \ov k_n}}\ .
\end{split}
\ee
In passing let us note that strong integrability of the models presented in \cite{Georgiou:2018gpe} with a deformation matrix of the form \eqref{lambda22} has been proven in \cite{Georgiou:2019plp}.

Finally, let us mention that, precisely as in section 3.2,  besides the $n$ equations of motion \eqref{semi-final} which can be put in the Lax form the remaining $N-n$ equations can be put in the form of covariantly free quantities.  Each of these equations implies the existence  of an infinite tower of conserved charges.

\subsection{Relaxing level conservation} 
In this section we consider the case where one drops the requirement of  level conservation at the vertices of the diagrams, that is we no longer impose the conditions $\tilde k_i=\hat k_i, \,\,\forall i$. 
In this case and in order to end up with an integrable theory one must demand that the fundamental theories one couples are all of the PCM-type and that the  coupling matrix is diagonal in the space of theories, namely that $(\tilde \l^{-1})_{ij}^{ab}=\l_i ^{-1}\,\d_{ij}\,\d^{ab}$.

To proceed we make in \eqref{semi-final} the following redefinitions $\tilde A_+^{(i)}=\sqrt{\hat k_i}{\cal A}_+^{(i)}$ and 
$\tilde A_-^{(i)}=\sqrt{\tilde k_i}{\cal A}_-^{(i)}$ to get
\be\label{fine-final}
\begin{split}
&\del_+{\cal A}_-^{(i)}-(\l_0^{(i)})^{-1}  \l_{i}^{-T}\del_-{\cal A}_+^{(i)}=(\l_0^{(i)})^{-1} [\l_{i}^{-T}{\cal A}_+^{(i)},{\cal A}_-^{(i)} ]\, ,\\
&\l_0^{(i)} \l_{i}^{-1}\del_+{\cal A}_-^{(i)}- \,\del_-{\cal A}_+^{(i)}=\l_0^{(i)}[{\cal A}_+^{(i)},\tilde \l_{i}^{-1} {\cal A}_-^{(i)} ],\,\qq \l_0^{(i)}=\sqrt{{\tilde k_i\ov \hat k_i}}.
\end{split}
\ee
Notice that this is precisely $n$ copies of the equations of motion (3.6) of the model presented in \cite{Georgiou:2017jfi}. The above equations of motion imply the existence of a Lax connection of the form \cite{Georgiou:2017jfi}
 \be
 \cL^{(i)}_\pm
=  {2 \,z_i \ov z_i \mp 1}  \check A^{(i)}_\pm \ ,\quad \check  A^{(i)}_+= {1-(\l_0^{(i)})^{-1} \l_1 \ov 1-\l_i^2} {\cal A}^{(i)}_+, \quad  \check A^{(i)}_-= {1-\l_0^{(i)} \l_i \ov 1-\l_i^2} {\cal A}^{(i)}_-
,\, \quad z_i\in\mathbb{C}\ .
\label{Laxpairs1}
\ee
As in section 3, one can straightforwardly  show by imitating the discussion below \eqref{drcurr} that the conserved charges obtained from the different Lax connections of \eqref{Laxpairs1} are in involution.  Let us, finally, note that,  as in section 3.2,  besides the $n$ equations of motion \eqref{fine-final} which can be put in the Lax form, the remaining $N-n$ equations can be put in the form of covariantly free combinations of currents.  Each of these equations implies the existence  of an infinite tower of conserved charges. 

A final comment is in order. Note that the cyclic $\l$-deformed models of \cite{Georgiou:2017oly} and \cite{Georgiou:2020eoo} belong to the class of the models of this subsection. In particular, the diagrams representing the aforementioned models are canonical polygons where each vertex is connected only to the adjacent ones.

\section{Conclusions}

In this work, we constructed the $\s$-model actions of a general class of integrable models. These models couple  N WZW models  with an arbitrary number of the following fundamental integrable theories, namely $n_1$ different copies of  the PCM, $n_2$ different copies of the YB model, both based on a group $G$, $n_3$ different copies of  the isotropic $\s$-model on the symmetric coset space $G/H$ and $n_4$ different copies of the YB model on the symmetric space $G/H$. The coupling is achieved by gauging 
the left global symmetry of the aforementioned fundamental integrable theories and connecting them with asymmetrically gauged WZW models. The action of the latter depends on both the gauge fields of the fundamental integrable theories which 
they connect.
In this way, webs of integrable theories are obtained. We show that a diagrammatic 
representation of these webs is possible. 
To each vertex of a diagram we assigned the matrix of one of the aforementioned fundamental integrable theories. Any two vertices may be connected with a number of lines with orientation  having  levels   $k_i$ with each of these lines   being associated to an asymmetrically gauged WZW model at an arbitrary level $k_i$.  Gauge invariance of the full action is translated to level conservation at the vertices. 
The virtue of this diagrammatic 
representation is that one can, at the back of the envelope, draw any diagram and directly write down from it the corresponding integrable theory.  A diagram which possesses all possible kinds of lines and vertices,  that is when each vertex is  connected to itself and to all other vertices in both directions,
corresponds to an integrable $\sigma$-model that depends on at least $n^2+1+n_2+n_4$ parameters, where $n$ is the total number of vertices/fundamental integrable models.
Next, we proved that the theories constructed  are indeed classically integrable by finding  the corresponding Lax pairs.

Subsequently, we considered two more general settings. In the first one, we focused on the case in which  the deformation matrix is not diagonal in the space of  the fundamental theories, in distinction to the theories of the 
previous sections.
In the second we examined the case in which,  although the deformation matrix is diagonal in the space of  the fundamental theories,  level conservation at the vertices is relaxed. In both cases we were able to prove integrability
only when all the deformation matrices are proportional to the identity in the group space, that is when only when all the theories we are coupling are all of the PCM-type.

There is a couple of  interesting questions remaining to be addressed. The first one concerns the quantum properties of the models presented. Although the $\beta$-functions of the couplings and the anomalous dimensions of the single currents can be straightforwardly deduced from the works \cite{Georgiou:2015nka,Georgiou:2016iom,Georgiou:2016zyo}
using perturbation theory around the conformal point  \footnote{ The  $\b$- functions and the single currents anomalous dimensions  of the models of section 2 are the same with those of the corresponding single $\l$-deformed models while the  $\b$- functions and the single currents anomalous dimensions of section 4.1 can be straightforwardly obtained from the analogous expressions in \cite{Georgiou:2018gpe} . Finally, the  $\b$- functions and the single current anomalous dimensions  of the models in section 4.2 are identical to the corresponding expressions of the two-level asymmetric construction of \cite{Georgiou:2016zyo,Georgiou:2017jfi} after identifying $\l_0$ of the latter paper with $\l_0^{(i)}$ of \eqref{fine-final}. The reason behind these identifications is that perturbation theory around the conformal points of the different theories is organised in  precisely the same way since it is of the current-current or parafermion-parafermion type. } the calculation of the exact anomalous dimensions of composite operators made from currents belonging to different WZW models, as well as those of the primary operators 
is certainly much more demanding since the result will be a non-trivial function of the couplings $\l_{ij}^{-1}$. The same holds for the three-point correlators involving currents and/or primary fields. Notice that for these calculations the 
methods developed in \cite{Georgiou:2019jcf} and \cite{Georgiou:2019aon} are more appropriate compared to the ones used in \cite{Georgiou:2015nka,Georgiou:2016iom,Georgiou:2016zyo} since for the theories of the present work we do not have the non-perturbative symmetries in the space of couplings which we had in 
the $\l$-deformed models with one or more parameters\cite{Kutasov:1989aw,Georgiou:2018hpd,Georgiou:2018gpe,Georgiou:2016zyo}. 

A second question concerns the Poisson-Lie T-dual theories of our models.
Given the relation between the $\l$- and $\eta$-deformations via Poisson-Lie T- duality and appropriate analytic continuations it is natural to wonder if  there are new integrable 
$\s$-models of the $\eta$-type to be constructed and which their relation will  be to the ones constructed based on the interpretation of integrable field theories as realisations of the affine Gaudin models\cite{Vicedo:2017cge,Delduc:2019bcl,Delduc:2018hty}.
In that respect, it would be important to study  the details of the algebraic and Hamiltonian structure of our theories. Finally, it would be very interesting to find the S-matrices of our theories and use the TBA equations in order to determine 
the corresponding mass gaps.

\section*{Acknowledgments}
We would like to thank Konstantinos Sfetsos for useful discussions and for an enjoyable and fruitful collaboration during the last years. 
This project has received funding from the Hellenic Foundation for Research and Innovation
(H.F.R.I.) and the General Secretariat for Research and Technology (G.S.R.T.), under grant
agreement No 15425. 

\appendix
\section{Comparison with other integrable models}

In order to be able to compare our theories with those  presented in    \cite{Georgiou:2018gpe}, notice that the  first term in  \eqref{sigma-fin} can be rewritten as an $N\times N$ matrix coupling the $N$ currents $J_+^{i=1,\dots,N}$ with the $N$ currents $J_-^{i=1,\dots,N}$, where $N$ is the {\it total} number of the WZW models. In doing so one can verify that the aforementioned $N\times N$ matrix  has zero determinant, it is not invertible and as a result it can never be written as the inverse of an $N\times N$ matrix of the form $\Lambda^{-T}-\cal D$ for some {\it regular }$\Lambda^{-T}$ and a diagonal $\cal D$ matrix as it is required by the models constructed in \cite{Georgiou:2018gpe} and in section 4.1 of \cite{Bassi:2019aaf}. To demonstrate this fact with an example consider the integrable theory of figure 1. In this case the first term in \eqref{sigma-fin} can be rewritten as 
\be\label{rewrite}
(J_{+\,11}^{(1)}, J_{+\,11}^{(2)},J_{+\,12}^{(1)},J_{+\,12}^{(2)},J_{+\,22}^{(1)},J_{+\,22}^{(2)},J_{+\,21}^{(1)},J_{+\,21}^{(2)})  \,\,\, .M_{8 \times 8}. \,\,\,
\left(       \begin{array}{c}
            J_{-\,11}^{(1)}   \\
            J_{-\,11}^{(2)} \\
             J_{-\,21}^{(1)} \\
              J_{-\,21}^{(2)}   \\
              J_{-\,22}^{(1)} \\
              J_{-\,22}^{(2)} \\
              J_{-\,12}^{(1)}\\
              J_{-\,12}^{(2)}\\
           \end{array}
         \right)\  ,
\ee
where the $8\times 8$ matrix $M_{8 \times 8}$ is
\be\label{M88}\small
\begin{split}
 &M_{8 \times 8}=\\
 &\left(      \begin{array}{cccccccc}
          \S_{11} (k_{11}^{(1)})^2&  \S_{11} k_{11}^{(1)} k_{11}^{(2)}& \S_{11} k_{11}^{(1)} k_{21}^{(1)}   & \S_{11} k_{11}^{(1)} k_{21}^{(2)}  &
           \S_{12} k_{11}^{(1)} k_{22}^{(1)}& \S_{12} k_{11}^{(1)} k_{22}^{(2)}& \S_{12} k_{11}^{(1)} k_{12}^{(1)}  & \S_{12} k_{11}^{(1)} k_{12}^{(2)}\\
            \S_{11} k_{11}^{(1)}k_{11}^{(2)}&  \S_{11} (k_{11}^{(2)})^2& \S_{11} k_{11}^{(2)} k_{21}^{(1)}   & \S_{11} k_{11}^{(2)} k_{21}^{(2)}  &
           \S_{12} k_{11}^{(2)} k_{22}^{(1)}& \S_{12} k_{11}^{(2)} k_{22}^{(2)}& \S_{12} k_{11}^{(2)} k_{12}^{(1)}  & \S_{12} k_{11}^{(2)} k_{12}^{(2)}\\
              \vdots  & \vdots  & \ddots & \vdots \\
              \vdots  & \vdots  & \ddots & \vdots \\
              \vdots  & \vdots  & \ddots & \vdots \\
           \end{array}  
         \right)\ , 
         \end{split}
\ee
where $  \S_{ij} =\Big({1\ov \l^{-1}-{\mathcal D}^T}\Big)_{ij},\,\,\,i,j=1,2$ and we have explicitly written only the first two lines. It is evident from \eqref{M88} that the second row of the matrix is equal to the first multiplied by 
$\nicefrac{k_{11}^{(2)}}{k_{11}^{(1)}}$. This implies that the determinant of $M_{8 \times 8}$ is zero and thus this matrix is not invertible, as discussed above. This result is a consequence of the fact that only certain linear combinations of the currents $J_\pm(g_{ij}^{(l_{ij})} )$ enter the first term of the action \eqref{sigma-fin}. Thus, we conclude that our models can not be written in any straightforward way in the form of those presented in  \cite{Georgiou:2018gpe}  and \cite{Bassi:2019aaf}.






\begin{thebibliography}{1}


\bibitem{Maldacena:1997re}
  J.~M.~Maldacena,
  {\it The Large N limit of superconformal field theories and supergravity},
  Int.\ J.\ Theor.\ Phys.\  {\bf 38} (1999) 1113 
  [Adv.\ Theor.\ Math.\ Phys.\  {\bf 2}, 231 (1998)],\hfill\break
  \href{https://arxiv.org/abs/hep-th/9711200}{arXiv: [hep-th/9711200]]}.

\bibitem{Staudacher:2004tk} 
  M.~Staudacher, {\it The Factorized S-matrix of CFT/AdS}, \hfill\break
   JHEP {\bf 0505} (2005) 054,
  \href{ https://arxiv.org/abs/hep-th/0412188}{arXiv: [hep-th/0412188]}.  

\bibitem{Ambjorn:2005wa} 
J.~Ambjorn, R.~A.~Janik and C.~Kristjansen,
{\it Wrapping interactions and a new source of corrections to the spin-chain/string duality},\hfill\break
Nucl. Phys. {\bf B736} (2006) 288,
\href{https://arxiv.org/abs/hep-th/0510171}{arXiv: [hep-th/0510171]}.

\bibitem{Gromov:2009tv} 
  N.~Gromov, V.~Kazakov and P.~Vieira,
   {\it Exact Spectrum of Anomalous Dimensions of Planar N=4 Supersymmetric Yang-Mills Theory},\hfill\break
  Phys. Rev. Lett.  {\bf 103} (2009) 131601,
  \href{https://arxiv.org/abs/0901.3753}{arXiv:0901.3753 [hep-th]}.


\bibitem{Beisert:2010jr}
  N.~Beisert {\it et al.},
  {\it Review of AdS/CFT Integrability: An Overview},\hfill\break
  Lett.\ Math.\ Phys.\  {\bf 99} (2012) 3,
  \href{https://arxiv.org/abs/1012.3982}{arXiv: 1012.3982 [hep-th]}.

\bibitem{Klimcik:2002zj}
C. Klim\v c\'\i k,
  {\it YB sigma models and dS/AdS T-duality},\hfill\break
  JHEP {\bf 0212} (2002) 051,
\href{http://arxiv.org/abs/hep-th/0210095}{hep-th/0210095.}

\bibitem{Klimcik:2008eq}
  C. Klim\v c\'\i k,
  {\it On integrability of the YB sigma-model},\hfill\break
  J. Math. Phys. {\bf 50} (2009) 043508,
  \href{http://arxiv.org/abs/0802.3518}{arXiv:0802.3518 [hep-th].}

 \bibitem{Klimcik:2014}
  C. Klim\v c\'\i k,
  {\it Integrability of the bi-Yang--Baxter sigma-model},
  Letters in Mathematical Physics {\bf 104} (2014) 1095,
    \href{http://arxiv.org/abs/1402.2105}{arXiv:1402.2105 [math-ph].}

 \bibitem{Delduc:2013fga}
  F.~Delduc, M.~Magro and B.~Vicedo,
{\it On classical $q$-deformations of integrable sigma-models},
  JHEP {\bf 1311} (2013) 192,
   \href{http://arxiv.org/abs/1308.3581}{arXiv:1308.3581 [hep-th].}

\bibitem{Delduc:2013qra}
  F.~Delduc, M.~Magro and B.~Vicedo,
{\it An integrable deformation of the $AdS_5 \times S^5$ superstring action},
  Phys. Rev. Lett. {\bf 112}, 051601,
     \href{http://arxiv.org/abs/1309.5850}{arXiv:1309.5850 [hep-th].}

\bibitem{Arutyunov:2013ega}
  G.~Arutyunov, R.~Borsato and S.~Frolov,
  {\it S-matrix for strings on $\eta$-deformed $AdS_{5} \times S^5$},
  JHEP {\bf 1404} (2014) 002,
 \href{http://arxiv.org/abs/1312.3542}{arXiv:1312.3542 [hep-th].}

\bibitem{Delduc:2014uaa}
F.~Delduc, M.~Magro and B.~Vicedo,
 {\it Integrable double deformation of the principal chiral model},
Nucl. Phys. B \textbf{891}, 312-321 (2015)
\href{https://arxiv.org/abs/1410.8066}{arXiv:1410.8066 [hep-th]}.


\bibitem{Delduc:2017fib}
F.~Delduc, B.~Hoare, T.~Kameyama and M.~Magro,
{\it Combining the bi-Yang-Baxter deformation, the Wess-Zumino term and TsT transformations in one integrable $\sigma$-model},
JHEP \textbf{10}, 212 (2017)
\href{https://arxiv.org/abs/1707.08371}{arXiv:1707.08371 [hep-th]}.



\bibitem{Klimcik:2016rov}
C. Klim\v c\'\i k,
  {\it Poisson--Lie T-duals of the bi-Yang--Baxter models},\hfill\break
  Phys. Lett.  {\bf B760} (2016) 345,
  \href{https://arxiv.org/abs/1606.03016}{arXiv:1606.03016 [hep-th]}.
  
   \bibitem{Klimcik:2015gba}
C. Klim\v c\'\i k,
  {\it $\eta$ and $\lambda$ deformations as ${\cal E}$-models},\hfill\break
   Nucl. Phys. {\bf B900} (2015) 259,
  \href{http://arxiv.org/abs/1508.05832}{arXiv:1508.05832 [hep-th].}

\bibitem{Klimcik:2017ken}
C.~Klimcik,
{\it Yang-Baxter $\sigma$-model with WZNW term as ${ \mathcal E}$-model},
Phys. Lett. B \textbf{772}, 725-730 (2017)
 \href{https://inspirehep.net/literature/1607828}{arXiv:1706.08912 [hep-th]}.


 \bibitem{Sfetsos:2013wia}
  K.~Sfetsos, {\it Integrable interpolations: From exact CFTs to non-Abelian T-duals},\hfill\break
  Nucl. Phys. {\bf B880} (2014) 225, \href{http://arxiv.org/abs/arXiv:1312.4560}{arXiv:1312.4560 [hep-th]}.



   
\bibitem{Georgiou:2016urf}
  G.~Georgiou and K.~Sfetsos,
  {\it A new class of integrable deformations of CFTs},\\
   JHEP {\bf 1703} (2017) 083,
  \href{https://arxiv.org/abs/1612.05012}{arXiv:1612.05012 [hep-th]}.


  \bibitem{Georgiou:2017jfi}
  G.~Georgiou and K.~Sfetsos,
  {\it Integrable flows between exact CFTs},\\
  JHEP {\bf 1711}  (2017) 078,
   \href{https://arxiv.org/abs/1707.05149}{arXiv:1707.05149 [hep-th]}.

\bibitem{Georgiou:2018hpd}
  G.~Georgiou and K.~Sfetsos,
  {\it Novel all loop actions of interacting CFTs: Construction, integrability and RG flows},
  Nucl.\ Phys. {\bf B937} (2018) 371,\href{https://arxiv.org/abs/1809.03522}{arXiv:1809.03522 [hep-th]]}.
 
\bibitem{Georgiou:2018gpe}
  G.~Georgiou and K.~Sfetsos,
  {\it The most general $\lambda$-deformation of CFTs and integrability},
    JHEP {\bf 1903} (2019) 094,
  \href{https://arxiv.org/abs/1812.04033} {arXiv:1812.04033 [hep-th]}.
 
\bibitem{Hollowood:2014rla} 
  T.J.~Hollowood, J.L.~Miramontes and D.M.~Schmidtt,
  {\it Integrable Deformations of Strings on Symmetric Spaces},
  JHEP {\bf 1411}, 009 (2014)
 \href{https://arxiv.org/abs/1407.2840}{[arXiv:1407.2840 [hep-th]}.
 
 \bibitem{Hollowood:2014qma}
  T.J.~Hollowood, J.L.~Miramontes and D.~Schmidtt,
{\it An Integrable Deformation of the $AdS_5 \times S^5$ Superstring},
J. Phys. {\bf A47} (2014) 49,  495402,
 \href{http://arxiv.org/abs/1409.1538}{arXiv:1409.1538 [hep-th]}.

\bibitem{Driezen:2019ykp}
  S.~Driezen, A.~Sevrin and D.~C.~Thompson,
  {\it Integrable asymmetric $\lambda$-deformations},
  JHEP {\bf 1904}, 094 (2019)
  \href{https://arxiv.org/abs/1902.04142}{arXiv:1902.04142 [hep-th]}.
  
   \bibitem{Witten:1983ar}
  E.~Witten,
  {\it Nonabelian Bosonization in Two-Dimensions},\hfill\break
  \href{https://link.springer.com/article/10.1007\%2FBF01215276}{Commun. Math. Phys.\  {\bf 92} (1984) 455.}   

\bibitem{Georgiou:2017oly}
  G.~Georgiou, K.~Sfetsos and K.~Siampos,
  {\it Double and cyclic $\lambda$-deformations and their canonical equivalents},
  Phys. Lett. {\bf B771}  (2017) 576,
   \href{https://arxiv.org/abs/1704.07834}{arXiv:1704.07834 [hep-th]}.



\bibitem{Georgiou:2015nka}
  G.~Georgiou, K.~Sfetsos and K.~Siampos,
  {\it All-loop anomalous dimensions in integrable $\lambda$-deformed $\sigma$-models},
  Nucl.\ Phys.\  {\bf B901} (2015) 40,
  \href{http://arxiv.org/abs/1509.02946}{arXiv:1509.02946 [hep-th].}



\bibitem{Georgiou:2016iom}
  G.~Georgiou, K.~Sfetsos and K.~Siampos,
  {\it All-loop correlators of integrable $\l$-deformed $\s$-models},
  Nucl.  Phys. {\bf B909} (2016) 360,
  \href{http://arxiv.org/abs/arXiv:1604.08212}{1604.08212 [hep-th].}


\bibitem{Georgiou:2016zyo}
  G.~Georgiou, K.~Sfetsos and K.~Siampos,
  {\it $\lambda$-deformations of left-right asymmetric CFTs}, Nucl. Phys. {\bf B914} (2017) 623,
\href{https://arxiv.org/abs/1610.05314}{arXiv:1610.05314 [hep-th]}.




    \bibitem{Itsios:2014lca}
  G.~Itsios, K.~Sfetsos and K.~Siampos,
  {\it The all-loop non-Abelian Thirring model and its RG flow},
  Phys.\ Lett.\  {\bf B733} (2014) 265,
  \href{http://arxiv.org/abs/1404.3748}{arXiv:1404.3748 [hep-th].}
  
       \bibitem{Georgiou:2017aei}
  G.~Georgiou, E.~Sagkrioti, K.~Sfetsos and K.~Siampos,
  {\it Quantum aspects of doubly deformed CFTs},
Nucl. Phys. {\bf B919} (2017) 504,
 \href{https://arxiv.org/abs/1703.00462}
  {arXiv:1703.00462 [hep-th]}.


 \bibitem{Sfetsos:2014jfa}
  K.~Sfetsos and K.~Siampos,
  {\it Gauged WZW-type theories and the all-loop anisotropic non-Abelian Thirring model},
  Nucl. Phys.  {\bf B885} (2014) 583,
  \href{http://arxiv.org/abs/arXiv:1405.7803}{arXiv:1405.7803 [hep-th].}


  \bibitem{Kutasov:1989aw}
  D.~Kutasov, {\it Duality Off the Critical Point in Two-dimensional Systems With Nonabelian Symmetries},
\href{http://www.sciencedirect.com/science/article/pii/0370269389913257}{Phys. Lett. {\bf B233} (1989) 369}.
  
   
\bibitem{Sagkrioti:2018rwg}
  E.~Sagkrioti, K.~Sfetsos and K.~Siampos,
  {\it RG flows for $\lambda$-deformed CFTs},\\
  Nucl.\ Phys.\ {\bf B930} (2018) 499,
    \href{https://arxiv.org/abs/1801.10174}{arXiv:1801.10174 [hep-th].}


    \bibitem{Kutasov:1989dt}
  D.~Kutasov,
  {\it String Theory and the Nonabelian Thirring Model},\\
 \href{http://www.sciencedirect.com/science/article/pii/0370269389912859}{Phys. Lett. {\bf B227} (1989) 68}.

  \bibitem{Gerganov:2000mt}
  B.~Gerganov, A.~LeClair and M.~Moriconi,
  {\it On the beta function for anisotropic current interactions in 2-D},
  Phys. Rev. Lett. {\bf 86} (2001) 4753,
 \href{http://arxiv.org/abs/hep-th/0011189}{hep-th/0011189}.


\bibitem{LeClair:2001yp}
  A.~LeClair,
  {\it Chiral stabilization of the renormalization group for flavor and color anisotropic current interactions},
  Phys.\ Lett.\ {\bf B519} (2001) 183,
  \href{https://arxiv.org/abs/hep-th/0105092v2}{hep-th/0105092}.

    \bibitem{Appadu:2015nfa}
  C. Appadu and T.J. Hollowood,
  {\it Beta function of k deformed ${\text AdS}_{5} \times S^5$ string theory},
  JHEP {\bf 1511} (2015) 095,
  \href{http://arxiv.org/abs/arXiv:1507.05420}{arXiv:1507.05420 [hep-th].}
  
\bibitem{Georgiou:2019jcf} 
  G.~Georgiou, P.~Panopoulos, E.~Sagkrioti and K.~Sfetsos,
  {\it Exact results from the geometry of couplings and the effective action},\hfill\break
   Nucl. Phys. {\bf B948} (2019) 114779, \href{https://arxiv.org/abs/1906.00984}{arXiv:1906.00984 [hep-th]}.
  
   \bibitem{Georgiou:2019aon} 
  G.~Georgiou and K.~Sfetsos,
 {\it Field theory and $\lambda$-deformations: Expanding around the identity},
  Nucl. Phys. {\bf B950}, 114855 (2020)
 \href{https://arxiv.org/abs/1910.01056}{[arXiv:1910.01056 [hep-th]}.
  



\bibitem{Zamolodchikov:1986gt}
  A.B. Zamolodchikov,
 {\it Irreversibility of the Flux of the Renormalization Group in a 2D Field Theory},
\href{http://www.jetpletters.ac.ru/ps/1413/article_21504.shtml}{JETP Lett.  {\bf 43} (1986) 730}.

\bibitem{Georgiou:2018vbb} 
  G.~Georgiou, P.~Panopoulos, E.~Sagkrioti, K.~Sfetsos and K.~Siampos,
  {\it The exact $C$-function in integrable $\lambda$-deformed theories},
  Phys.\ Lett.\ B {\bf 782}, 613 (2018)
   \href{https://arxiv.org/abs/1805.03731} {[arXiv:1805.03731 [hep-th]}.
  
\bibitem{Sagkrioti:2018abh} 
  E.~Sagkrioti, K.~Sfetsos and K.~Siampos,
  {\it Weyl anomaly and the $C$-function in $\lambda$-deformed CFTs},
  Nucl.\ Phys.\ B {\bf 938}, 426 (2019)
  \href{https://arxiv.org/abs/1810.04189}{[arXiv:1810.04189 [hep-th]}.

 \bibitem{Georgiou:2019nbz} 
  G.~Georgiou, E.~Sagkrioti, K.~Sfetsos and K.~Siampos,
  {\it An exact symmetry in $\lambda$-deformed CFTs},
  JHEP {\bf 2001}, 083 (2020)
  \href{https://arxiv.org/abs/1911.02027}{[arXiv:1911.02027 [hep-th]}.
  
\bibitem{Georgiou:2020bpx}
G.~Georgiou, K.~Sfetsos and K.~Siampos,
{\it A free field perspective of $\lambda$-deformed coset CFT's},
\href{https://inspirehep.net/literature/1792132}{[arXiv:2004.10216 [hep-th]}.
  

  
        \bibitem{Hoare:2019mcc} 
  B.~Hoare, N.~Levine and A.~A.~Tseytlin,
  {\it Integrable sigma models and 2-loop RG flow},
  JHEP {\bf 1912}, 146 (2019)
  \href{https://arxiv.org/abs/1910.00397}{[arXiv:1910.00397 [hep-th]}.
    



\bibitem{Sfetsos:2017sep}
  K.~Sfetsos and K.~Siampos,
  {\it Integrable deformations of the $G_{k_1} \times G_{k_2}/G_{k_1+k_2}$ coset CFTs},
  Nucl. Phys. {\bf B927} (2018) 124,
  \href{https://arxiv.org/abs/1710.02515}{arXiv:1710.02515  [hep-th]}.


\bibitem{Balog:1993es}
  J.~Balog, P.~Forgacs, Z.~Horvath and L.~Palla,
  {\it A New family of $SU(2)$ symmetric integrable sigma models,}
  Phys. Lett. {\bf B324} (1994) 403,
  \href{http://arxiv.org/abs/hep-th/9307030}{hep-th/9307030.}
  
\bibitem{Georgiou:2020eoo}
G.~Georgiou, G.~Pappas and K.~Sfetsos,
[arXiv:2005.02414 [hep-th]].

\bibitem{Vicedo:2015pna}
  B.~Vicedo,
  {\it Deformed integrable $\sigma$-models, classical $R$-matrices and classical exchange algebra on Drinfel'd doubles},
  \hfill\break
  J. Phys. A: Math. Theor. {\bf 48} (2015) 355203,
 \href{http://arxiv.org/abs/1504.06303}{arXiv:1504.06303 [hep-th].}


\bibitem{Hoare:2015gda}
  B.~Hoare and A.A.~Tseytlin,
  {\it On integrable deformations of superstring sigma models related to $AdS_n \times S^n$ supercosets},\hfill\break
  {Nucl. Phys. {\bf B897} (2015) 448},
  \href{http://arxiv.org/abs/1504.07213}{arXiv:1504.07213 [hep-th].}


 

        \bibitem{Sfetsos:2015nya}
  K.~Sfetsos, K.~Siampos and D.C.~Thompson,
 {\it Generalised integrable $\lambda$- and $\eta$-deformations and their relation},\hfill\break
  Nucl. Phys. {\bf B899} (2015) 489,
  \href{http://arxiv.org/abs/1506.05784}{arXiv:1506.05784 [hep-th].}

\bibitem{Demulder:2020dlo}
S.~Demulder, F.~Hassler, G.~Piccinini and D.~C.~Thompson,
{\it Integrable deformation of $\mathbb{CP}^n$ and generalised K\"ahler geometry},
JHEP \textbf{10}, 086 (2020)
\href{https://arxiv.org/abs/2002.11144}{arXiv:2002.11144 [hep-th]}.

   
\bibitem{Hoare:2018ebg}
  B.~Hoare and F.K.~Seibold,
 {\it Poisson-Lie duals of the $\eta$-deformed $\mathrm{AdS}_2 \times \mathrm{S}^2 \times \mathrm{T}^6$ superstring},
  JHEP {\bf 1808} (2018) 107,
 \href{https://arxiv.org/abs/1807.04608}{arXiv:1807.04608 [hep-th]}.
  
\bibitem{Driezen:2018glg}
  S.~Driezen, A.~Sevrin and D. C.~Thompson,
  {\it D-branes in $\lambda$-deformations},\hfill\break
  JHEP {\bf 1809} (2018) 015,
 \href{https://arxiv.org/abs/1806.10712}{arXiv:1806.10712 [hep-th]}.

\bibitem{KS95a}{C. Klim\v c\'\i k and P. \v Severa, {\it Dual non-Abelian duality and the Drinfeld double},\\
Phys. Lett. {\bf B351}
(1995) 455, \href{http://arxiv.org/abs/hep-th/9502122}{hep-th/9502122}.}

\bibitem{Sfetsos:1999zm}
  K.~Sfetsos,
  {\it Duality invariant class of two-dimensional field theories},\hfill\break
  Nucl. Phys. {\bf B561} (1999) 316,
  \href{https://arxiv.org/abs/hep-th/9904188}{[hep-th/9904188]}.
  
\bibitem{Witten:1991mm}
  E. Witten,
  {\it On Holomorphic factorization of WZW and coset models},\hfill\break
\href{http://link.springer.com/article/10.1007\%2FBF02099196} {Commun. Math. Phys.  {\bf 144} (1992) 189}.

  \bibitem{Bowcock} P. Bowcock, {\it  Canonical Quantization of the Gauged Wess-Zumino Model},\hfill\break \href{https://www.sciencedirect.com/science/article/pii/0550321389903878}{Nucl. Phys. {\bf B316}(1989) 80}

\bibitem{Georgiou:2019plp}
G.~Georgiou, K.~Sfetsos and K.~Siampos,
 {\it Strong integrability of $\lambda$-deformed models},
Nucl. Phys. B \textbf{952}, 114923 (2020)
\href{https://arxiv.org/abs/1911.07859}
{arXiv:1911.07859 [hep-th]}.

\bibitem{Evans:1999mj}
J.~M.~Evans, M.~Hassan, N.~J.~MacKay and A.~J.~Mountain,
 {\it Local conserved charges in principal chiral models},
Nucl. Phys. B \textbf{561}, 385-412 (1999)
\href{https://arxiv.org/abs/hep-th/9902008v2}
{arXiv:hep-th/9902008 [hep-th]}.

\bibitem{Bassi:2019aaf} 
  C.~Bassi and S.~Lacroix,
   {\it Integrable deformations of coupled $\sigma$-models},
  \href{https://arxiv.org/abs/1912.06157}{arXiv:1912.06157 [hep-th]}.
  
  \bibitem{Georgiou} 
  G.~Georgiou, work in progress.

\bibitem{Vicedo:2017cge}
B.~Vicedo,
 {\it  On integrable field theories as dihedral affine Gaudin models},
\href{https://arxiv.org/abs/1701.04856}{arXiv:1701.04856 [hep-th]}.

\bibitem{Delduc:2019bcl} 
  F.~Delduc, S.~Lacroix, M.~Magro and B.~Vicedo,
 { \it Assembling integrable $\sigma$-models as affine Gaudin models},
  JHEP {\bf 1906}, 017 (2019)
 \href{https://arxiv.org/abs/1903.00368}{[arXiv:1903.00368 [hep-th]}.

\bibitem{Costello:2019tri}
K.~Costello and M.~Yamazaki,
{ \it Gauge Theory And Integrability, III},
\href{https://arxiv.org/abs/1908.02289}{arXiv:1908.02289 [hep-th]}.

\bibitem{Delduc:2018hty}
  F.~Delduc, S.~Lacroix, M.~Magro and B.~Vicedo, {\it Integrable coupled sigma-models},
   Phys. Rev. Lett. {\bf 122} (2019) no.4, 041601,
  \href{https://arxiv.org/abs/1811.12316}{ arXiv:1811.12316 [hep-th]}.


\end{thebibliography}
\end{document}